\documentclass[12pt]{article}
\newcommand{\Comment}[1]{{}}
\usepackage[textwidth = 450 pt, textheight = 630 pt]{geometry}
\usepackage[parfill]{parskip}
\usepackage{amssymb,euscript,amsmath,amsfonts}
\usepackage{tikz,tikz-cd,url}
\usetikzlibrary{decorations.pathreplacing,calligraphy}
\usepackage{xcolor}
\usepackage{graphicx}
\usepackage{color}
\usepackage{bbm}
\usepackage{tensor}
\usepackage{slashed}
\usepackage{bigints}
\usepackage{mathrsfs}

\newcommand{\bb}[1]{\mathbb{#1}}

\renewcommand{\frak}[1]{\mathfrak{#1}}
\renewcommand{\cal}[1]{\mathcal{#1}}

\newcommand{\inv}[1]{\frac{1}{#1}}
\newcommand{\calz}{\cal{Z}}
\newcommand{\bcalz}{\Bar{\calz}}
\newcommand{\brac}[1]{\left(#1\right)}

\newcommand{\mn}{\mu\nu}

\renewcommand{\dag}{\dagger}

\newcommand{\bpsi}{\Bar{\psi}}
\newcommand{\hpsi}{\hat{\psi}}

\newcommand{\balpha}{\Bar{\alpha}}

\newcommand{\bxi}{\Bar{\xi}}

\newcommand{\hcalz}{\hat{\calz}}
\newcommand{\halpha}{\hat{\alpha}}
\newcommand{\hX}{\hat{X}}
\newcommand{\hY}{\hat{Y}}

\newcommand{\hxi}{\hat{\xi}}

\newcommand{\cE}{\cal{E}}

\allowdisplaybreaks

\definecolor{MyDarkBlue}{rgb}{0.15,0.15,0.45}
\usepackage[linktocpage=true]{hyperref}
\hypersetup{
colorlinks=true,
citecolor=MyDarkBlue,
linkcolor=MyDarkBlue,
urlcolor=MyDarkBlue,
pdfauthor={Neil Lambert },
pdftitle={ },
pdfsubject={hep-th}
}

\usepackage{hypernat}
\usepackage{physics}
\usepackage{accents}
\usepackage{bm}
\usepackage[sorting=none, style=numeric-comp]{biblatex}
\begin{document}

   \vspace{1.8truecm}

 \centerline{\Huge  {Reciprocal Non-Relativistic Decoupling}}
 \centerline{\Huge{Limits of String Theory and M-Theory}} \vskip 12pt

\centerline{\LARGE \bf {\sc  }} \vspace{2truecm} \thispagestyle{empty} \centerline{
    {\large {{\sc Neil~Lambert}}}\footnote{E-mail address: \href{mailto:neil.lambert@kcl.ac.uk}{\tt neil.lambert@kcl.ac.uk} } and  {\large {{\sc Joseph~Smith}}}\footnote{E-mail address: \href{mailto:joseph.m.smith@kcl.ac.uk}{\tt joseph.m.smith@kcl.ac.uk} }}

\vspace{1cm}
\centerline{{\it Department of Mathematics}}
\centerline{{\it King's College London }} 
\centerline{{\it The Strand }} 
\centerline{{\it  WC2R 2LS, UK}} 

\vspace{1.0truecm}

\thispagestyle{empty}

\centerline{\sc Abstract}
\vspace{0.4truecm}
\begin{center}
\begin{minipage}[c]{360pt}{
    \noindent}

It has recently been realised that there are supersymmetry-preserving non-relativistic decoupling limits associated with each half-BPS object in String Theory and M-Theory. We argue that, given a $p$-brane and a $q$-brane for which there is a quarter-BPS intersecting configuration, the $p$-brane decoupling limit of the $q$-brane's worldvolume QFT is necessarily equal to the reciprocal $q$-brane decoupling limit of the $p$-brane's worldvolume theory. This is explicitly shown for the cases of the D0-D4 and D1-D3 systems. As an application of this idea, we use the reciprocal pair of limits in the M2-M5 system to describe the dynamics of interacting self-dual strings in the six-dimensional $\mathcal{N}=(2,0)$ theory. We also discuss the limits of the dual gravitational theories and, using symmetries, argue that the duality is maintained by the limits.

\end{minipage}
\end{center}

\newpage
\tableofcontents

\section{Introduction}\label{sect: Intro}

In the last few years there has been a renewed interest in limits of String Theory and M-theory \cite{Harmark:2019upf, Bergshoeff:2019pij, Harmark:2020vll, Bergshoeff:2021tfn,  Bidussi:2021ujm, Blair:2021waq, Blair:2023noj, Fontanella:2024rvn, Hartong:2024ydv}. A topic of considerable intrigue is decoupling limits associated with half-BPS extended objects, which are generalisations of the familiar Matrix Theory limit \cite{Banks:1996vh, Susskind:1997cw} of type IIA String Theory generated using dualities and higher-dimensional uplifts. In order to define these limits one chooses an alignment of the given brane in the bulk spacetime. The limits then act as non-Lorentzian limits of the theory, deforming the metric structure of spacetime into generalised Newton-Cartan (NC) geometries in which the relativistic metric is split into a `longitudinal' piece aligned with the brane and a 'transverse' piece comprising the remaining directions. The limits of interest in this work will be those associated with D$p$-branes, which we refer to as D$p$NC limits, those associated with the fundamental string, which we refer to as SNC limits, and those associated with M-theory branes, which we refer to as M$p$NC limits.

In recent work we discussed the application of various NC limits associated to branes in String Theory and M-theory \cite{Lambert:2024uue, Lambert:2024yjk}. In particular, we argued that taking a D$q$NC Limit of a D$p$-brane gauge theory yields a non-relativistic field theory that describes the intersection of a D$p$-brane with a D$q$-brane (and similarly  an M2NC limit of the M2-brane conformal field theory (CFT) leads to a non-relativistic Chern-Simons-Matter theory describing the intersection of two stacks of M2-branes). This interpretation has an immediate consequence: if the D$q$NC limit of the D$p$-brane gauge theory describes the intersection of D$q$-branes with D$p$-branes, then the D$p$NC limit of the D$q$-brane theory must also describe the same intersection. These two apparently different limits must therefore land on the same dynamical system. This is somewhat remarkable, as for generic cases the two live in different dimensions. This disparity is resolved by noting that the limits impose constraints on the system that reduce the configuration space of the theory to a moduli space, leading to effective lower-dimensional systems. As we will see below, in the case of D-branes the equivalence then arises through a manifestation of the Nahm transform, with an isomorphism between the resulting moduli spaces arising from both limits.

An interesting playground in which limits of this sort can be studied is M-theory. Since there are two types of BPS branes in M-theory, M2-branes and M5-branes, we would expect that there is a consistent supersymmetry-preserving decoupling limit associated with each. We will refer to these as the M2NC and M5NC limits respectively. The M2NC limit of the Bosonic sector of eleven-dimensional supergravity was taken in \cite{Blair:2021waq}, with the supersymmetric completion constructed in \cite{Bergshoeff:2024nin}. The M5NC limit of the supergravity theory has not been constructed, though is believed to exist. In order to find half-BPS limits of the worldvolume field theories we must therefore consider the quarter-BPS brane configurations with dynamical intersections. Excluding the M2-M2 intersection already studied in \cite{Lambert:2024uue}, there are two such set-ups:
M5-M5 intersections
\begin{align} \label{eq: M5-M5 intersection}
\begin{array}{rrrrrrrrr}
    M5: & 0 & 1 & 2 & 3 & 4 & 5 & &\\
    M5:& 0 & 1 & 2 &3 & &  & 6 & 7 \ \ , \\
\end{array}
\end{align}
and M2-M5 intersections
\begin{align}
\begin{array}{rrrrrrrr}
    M2: & 0 & 1 & 2 & & & & \\
    M5:& 0 & 1 & &3 &4 & 5 & 6 \ \ . \\
\end{array}
\end{align}
As in \cite{Lambert:2024yjk}, we can then take decoupling limits by considering one of the branes in the configuration as defining either the M2NC or M5NC limits, and consider the low-energy worldvolume QFT on the other in this limit. 

As the first set-up is symmetric under the interchange of the branes there is only one limit that can be taken. The theory describing the intersection is therefore the M5NC limit of the six-dimensional $\cal{N}=(2,0)$ theory. This is somewhat more exotic than the previously studied cases, as a complete description of the field theory is not known and we therefore cannot directly take the limit. However, we can make progress by compactifying \eqref{eq: M5-M5 intersection} on a spacelike torus common to both branes to reach the configuration
\begin{align}
\begin{array}{rrrrrrr}
    D3: & 0 & 1 & 2 & 3 &  &\\
    D3: & 0 & 1 &  & & 4 & 5  \ \ , \\
\end{array}
\end{align}
in type IIB string theory. The limit associated with this configuration is the D3NC limit of $\cal{N}=4$ SYM \cite{Bershadsky:1995vm, Lambert:2024yjk}, which localises us onto solutions of Hitchin's equations in the directions transverse to the D3NC limit. As this set of equations is solely associated with the relative transverse directions, it cannot be changed by dimensional reduction along directions longitudinal to both branes. This means that we can lift the discussion back to the original M5-M5NC case, with the result being that the limit of the $\cal{N}=(2,0)$ theory is a four-dimensional theory in the longitudinal coordinates determined by a solution of Hitchin's equations in the transverse coordinates. After a topological twist we may take these coordinates, which were originally those of a torus, to now parameterise an arbitrary Riemann surface. It therefore seems natural to identify this limit with the class-$\cal{S}$ construction \cite{Gaiotto:2009hg,Gaiotto:2009we} of four-dimensional SCFTs, and the resulting M5NC field theory with the theory on the Coulomb branch.

The second example is the M2-M5 intersection. From the perspective of the reciprocal pair of limits this can be viewed as either the M2NC limit of the six-dimensional $\cal{N}=(2,0)$ theory on the M5-branes, or as the M5NC limit of the BLG or ABJM theory on the M2-branes. As there is no Lagrangian description of the $\cal{N}=(2,0)$ theory we cannot directly carry out the first limit. We do, however, have a Lagrangian description of the latter, and our physical picture of the limits tells us that the theories must be equivalent. Amazingly, even though we do not understand the six-dimensional theory we can say quantitative things about certain sectors of its dynamics. In this case, the decoupling limit isolates the self-dual strings in the theory that arise as the endpoints of M2-branes on the M5-branes \cite{Howe:1997ue,Haghighat:2013gba}.

In our previous work we also discussed the corresponding limit of the associated supergravity solutions. These contain AdS-factors of a smaller dimension that the original gravity-dual, reflecting the restriction of the dynamics to the intersection. There is therefore hope that the AdS/CFT correspondence remains true after the appropriate limits. This could in principle give new examples of AdS/CFT pairs, but outside the usual context. Furthermore, it would seem that these new duals are in some sense simpler than the traditional AdS/CFT pairs, involving only a subsector of the dynamical degrees of freedom of the original theory. In this paper we will generalise this to discussion to include a reciprocal equivalence in the gravitational duals. Since the two equivalent field theories are found by taking limits of theories with gravitational duals, if this duality is not broken by the limit then the limits of the gravitational theories must also be equivalent. We will argue here that the gravitational duals are warped products of Minkowski space and various Euclidean planes which, within the context of Newton-Cartan geometry, nevertheless posses the same symmetries as an AdS background within standard geometry, possibly broken by a non-trivial dilaton. Our gravitational solutions here are somewhat different to the ones we considered in\cite{Lambert:2024uue,Lambert:2024yjk} where we found gravitational duals with AdS factors. The difference arises from whether or not one takes the number of branes to scale with $c$; we will comment more on this distinction later. Thus, for any given brane intersection (with $p\ne q$) we find four possible descriptions - two field theories and their gravitational duals - depending on which Newton-Cartan limit we take. They therefore present an interesting opportunity to probe how gauge/gravity duality works in a non-AdS environment. We note that while we shall present evidence that the duality is preserved by the limit, it is possible that the AdS/CFT correspondence is not somehow broken by this procedure. This would arguably be more interesting, as by breaking the duality we would gain some insight into what makes  AdS/CFT work.

Our goal in this work is to provide evidence that this interpretation of decoupling limits is indeed accurate. We will use the well-studied examples of the D0-D4 and D1-D3 systems as a testing ground for the reciprocal limits, and show that in both cases the two field theories obtained are exactly equal. We can then apply the general framework to the case of the M2-M5 system, giving us a Lagrangian description of interacting self-dual strings within the six-dimensional $\cal{N}=(2,0)$ theory. In all three cases we will also discuss the corresponding Newton-Cartan limits of the various gravitational duals. By matching symmetries, we will argue that gauge-gravity duality is maintained throughout the limit, as predicted.

The rest of this paper is organised as follows. In section \ref{sect: General} we will outline how symmetries work in a special class of gravitational backgrounds in Newton-Cartan gravities which will be useful in the following analysis. In \ref{sect: d0-d4} we consider the case of intersecting D0-branes and D4-branes, both from the field theory point of view and gravity duals arguing that they lead to conformal quantum mechanics with NC gravitational duals. In section \ref{sect: d1-d3} we examine the D1-D3 system in both field theory and gravity. Here we find that conformal symmetry is broken as the dynamics takes place on the Coulomb branch of the gauge theory. This is reflected in the gravitational dual by the presence of a non-trivial dilaton. In  section \ref{sect: m2-m5} we will explore the NC limits associated with intersecting M2-branes and M5-branes. Here we again find a conformal system with a gravitational dual that we associate with the low-energy dynamics of a self-dual string living on the intersection. Finally we conclude with our conclusions in section \ref{sect: conclusion}. We also include several appendices with further details of our calculations.

While we were completing this paper we received \cite{Blair:2024aqz}. This paper presents a unifying view   of non-relativistic decoupling limits and matrix theories ({\it aka} Yang-Mills theories associated to D$p$-branes). Our  results  fit  very naturally into this framework and our discussion is complimentary to theirs. In particular our limits correspond to the double scaling limits of  \cite{Blair:2024aqz} and reciprocity corresponds to the statement that the two limits commute.

\section {\texorpdfstring{Symmetries in $p$-brane Newton-Cartan Geometry}{Symmetries in p-brane Newton-Cartan Geometry}}\label{sect: General}

Before looking at the examples let us first make some general comments  of symmetries in $p$-brane Newton-Cartan (NC) geometries. To reach these from Lorentzian geometries we consider a split of the relativistic veilbein into two groupings
\begin{equation}
    \hat{e}^{\hat a} = \{ c \tau^A , c^{-\lambda/2} e^a \} \ ,
\end{equation}
where $A = 0, 1 , \ldots , p$, and $a$ parameterises the remainder of the indices. Throughout this paper $c$ is a dimensionless constant that we ultimately take to infinity and here $\lambda\ge0$ is a constant depending on the type of NC limit ($\lambda=0$ for SNC, $\lambda=1$ for M2NC, $\lambda=2$ for D$p$NC, and $\lambda = 4$ for M5NC). Though both $\tau^A$ and $e^a$ can in principle have additional sub-lieading $c$-dependence this will not change the dynamics of the limits we will consider and we will proceed as if they are $c$-independent. When the concept is well-defined we call the dimensions associated with $\tau^A$ `large' and those associated with $e^a$ `small'. A similar split occurs in the dual orthonormal basis. Defining the contractions
\begin{subequations}
\begin{align}
    \cal{T}_{\mn} &= \eta_{AB} \tau^A_{\mu} \tau^B_{\nu} \ , \\
    \cal{E}_{\mn} &= \delta_{ab} e^a_{\mu} e^b_{\nu} \ ,
\end{align}
\end{subequations}
and analogously for the orthonormal basis, we find that the metric and cometric decompose into
\begin{subequations}
\begin{align}\label{gEC}
    g_{\mn} &= c^2 {\cal T}_{\mn} + c^{-\lambda} {\cal E}_{\mn} \ , \\
    g^{\mn} &= c^\lambda {\cal E}^{\mn} + c^{-2}{\cal T}^{\mn} \ .
\end{align}  
\end{subequations}

For finite $c$ this decomposition is invariant under $SO(1,D-1)$ Lorentz transformations of the veilbeins, which now contain factors of $c$. When the $c\to \infty$ limit is taken the surviving transformations are those that do not diverge; this breaks the Lorentz group to $SO(1,p)$ transformations that act on $\tau^I$, $SO(D-p-1)$ transformations that act on $e^i$, and Galilean boosts that act on $e^a_{\mu}$ and $\tau^{\mu}_A$ as
\begin{subequations}
\begin{align}
    e^a_{\mu} &\to e^a_{\mu} - V^a_A \tau_{\mu}^A \ ,  \\
    \tau^{\mu}_A &\to \tau^{\mu}_A + V_A^a e_a^{\mu} \ ,
\end{align}
\end{subequations}
in terms of some coefficients $V^a_A$ that parameterise the transformation. However, unlike in the Lorentzian case these transformations act on the contracted tensor fields $\cal{E}_{\mn}$ and $\cal{T}^{\mn}$ as
\begin{align}\label{eq: Gboost}
    {\cal E}_{\mn} &\to {\cal E}_{\mn} - 2 V_{aA} e^a_{(\mu} \tau^A_{\nu)} \ , \\
    \cal{T}^{\mn} & \to \cal{T}^{\mn} + 2 V^{aA} e_a^{(\mu} \tau_A^{\nu)} \ .
\end{align}
As these are symmetries of the description these fields are only well-defined up to an equivalence-class under the transformations \eqref{eq: Gboost}; the physical fields of the description are $\cal{T}_{\mn}$ and $\cal{E}^{\mn}$. In effect this means that, at least to first order,  any cross terms in ${\cal E}_{\mn}$  between coordinates that lie in ${\cal T}_{\mn}$ and ones which lie in ${\cal E}_{\mn}$ can be gauged away. An intrinsic definition of the geometry is given by taking $(\cal{T}_{\mn}, \cal{E}^{\mn})$ to be degenerate symmetric tensor fields such that, on a manifold of dimension $D$,
\begin{itemize}
    \item $\cal{T}_{\mn}$ has a single negative eigenvalue, $p$ positive eigenvalues, and a $(D-p-1)$-dimensional subspace with vanishing eigenvalues.
    \item $\cal{E}^{\mn}$ has $D-p-1$ positive eigenvalues and a $(p+1)$-dimensional subspace with vanishing eigenvalues.
    \item The subspaces of non-vanishing eigenvalues of each field are disjoint, \textit{i.e.} $\cal{T}_{\mu\rho} \cal{E}^{\rho\nu} = 0$.
\end{itemize} 
We can then define the projective inverses $(\cal{T}^{\mn}, \cal{E}_{\mn})$ by requiring that the relations
\begin{subequations}
\begin{align}
    \cal{T}_{\mu\rho} \cal{T}^{\rho\nu} + \cal{E}_{\mu\rho} \cal{E}^{\rho\nu} &= \delta^{\nu}_{\mu} \ , \\
    \cal{T}^{\mu\rho} \cal{E}_{\rho\nu} &= 0 \ ,
\end{align}
\end{subequations}
are satisfied, which are invariant under the transformations \eqref{eq: Gboost}.

In this paper we consider metrics that arise from intersecting brane configurations which take the form, in the near horizon limit, 
\begin{align} \nonumber
g =	\,&c^2\left[\left(\frac{R}{|Y|}\right)^{2\alpha}\eta_{\mu\nu }d\sigma^\mu d\sigma^\nu+\left(\frac{R}{|Y|}\right)^{2\beta}dX\cdot dX \right] \\ \label{eq: NCg} 
& + c^{-\lambda} \left[\left(\frac{R}{|Y|}\right)^{2\gamma}dx\cdot dx    + \left(\frac{R}{|Y|}\right)^{2\delta}dY\cdot dY \right] \ .   
\end{align}
Here the large directions parameterize a product of Minkowski space with a  plane,  whereas the small directions are products of planes. These come with various warp factors that are simple powers of $|Y|/R$ for some constant $R$. Without loss of generality we assume that $\alpha\ne\beta$ so that the plane parameterised by the $X$-coordinates is distinct from the spatial plane of Minkowski space. We do, however, allow for the possibility of there being no $X$ or $x$ coordinates, with $\cal{T}$ or $\cal{E}$ reducing to the metric of a single plane up to a conformal factor.

Ultimately, one takes the limit $c\to\infty$ where (\ref{eq: NCg}) no longer makes sense\footnote{ Crucially, the limit does make sense in the supergravity action and one obtains a consistent dynamical system including the remaining supergravity fields as $c\to\infty$.}. On a formal level we must therefore think of ${\cal T}_{\mn}$ and ${\cal E}_{\mn}$ as independent components of the NC structure, with
\begin{subequations}
\begin{align}\label{eq: TE}
    {\cal T} &= \left(\frac{R}{|Y|}\right)^{2\alpha}\eta_{\mu\nu }d\sigma^\mu d\sigma^\nu+\left(\frac{R}{|Y|}\right)^{2\beta}dX\cdot dX \nonumber\\
    {\cal E} & = \left(\frac{R}{|Y|}\right)^{2\gamma}dx\cdot dx    + \left(\frac{R}{|Y|}\right)^{2\delta}dY\cdot dY  \ ,
\end{align}
\end{subequations}
rather than combining them into a Lorentzian metric as we have done above. Nevertheless, we find it more convenient to write the metric explicitly in the form of (\ref{eq: NCg}), though it is important to bear in mind that this is just a notational choice; we are really talking about  ${\cal T}$ and ${\cal E}$ as separate entities that arise in the $c\to\infty$ limit.

Let us consider the symmetries of such spacetimes within the context of NC gravity. For the supersymmetry-preserving limits of interest here there is an additional local scaling symmetry
\begin{subequations} \label{eq: ncw general transformation}
\begin{align}
    \cal{T}_{\mn} &\to \Lambda^2 \cal{T}_{\mn} \ , \\
    \cal{E}_{\mn} &\to \Lambda^{-\lambda} \cal{E}_{\mn} \ ,
\end{align}
\end{subequations}
where $\Lambda$ is any function of the coordinates, that we refer to as a Newton-Cartan-Weyl (NCW) transformation\footnote{This symmetry is not present for generic NC limits: see \cite{Bergshoeff:2024ipq}.}. Informally, we may think of this as a rescaling of $c\to {\Lambda}c$ before taking the limit. To account for this, when $\lambda\ne0$\footnote{The only case of interest with $\lambda=0$ is the SNC limit. In this case, the dilaton also transforms under this symmetry and so we can fix the symmetry by requiring that it vanishes.} we can without loss of generality fix this additional scale symmetry by mapping the metric  to the form
\begin{align} \nonumber
    g' = \, &	c^2\left[\left(\frac{R}{|Y|}\right)^{2\alpha'}\eta_{\mu\nu }d\sigma^\mu d\sigma^\nu+\left(\frac{R}{|Y|}\right)^{2\beta'}dX\cdot dX \right] \\ \label{eq: NCg1}
    &+ c^{-\lambda} \left[\left(\frac{R}{|Y|}\right)^{2\gamma'}dx\cdot dx    + \left(\frac{R}{|Y|}\right)^{2}dY\cdot dY \right] \ ,
\end{align}
where $\alpha'=\alpha-4\lambda^{-1}(1-\delta)$, $\beta'=\beta-4\lambda^{-1}(1-\delta)$ and $\gamma'=\gamma+1-\delta$. Once this scale symmetry is fixed the symmetries correspond to transformations for which $\cal{T}_{\mn}$ and $\cal{E}^{\mn}$ are invariant. It will also be convenient to think of this in terms of the metric \eqref{eq: NCg1}, where the symmetries correspond to Killing vectors (as is the case in usual Riemannian geometry) up to transformations of $\cal{E}_{\mn}$ that can be undone using the Galilean boost symmetry (\ref{eq: Gboost}). 

The obvious symmetries of \eqref{eq: NCg1} are the Poincar\'e symmetry group of the Minkowskian coordinates $\sigma^\mu$, the Euclidean symmetry groups of the planes with coordinates $x^i$ and $X^I$, and the rotational symmetry group of the $Y^M$ plane. For this class of solution we also always find a simple scaling symmetry
\begin{subequations} \label{eq: rigidC}
\begin{align}
    Y^M&\to \Omega Y^M\nonumber\\
    x^i&\to \Omega^{\gamma'}x^i\nonumber\\
    X^I&\to \Omega^{\beta'} X^I\nonumber\\
    \sigma^\mu &\to \Omega^{\alpha'} \sigma^\mu\ ,
\end{align}
\end{subequations}
where $\Omega$ is a constant, and we may wonder whether there are circumstances in which this enhances to a larger conformal group. Let us first assume that $\alpha'\neq0$. We can then consider the combination of a conformal transformation
\begin{subequations}
 \begin{align}
    \sigma^\mu &\to \sigma'^\mu \ , \\
    \eta_{\mu\nu }d\sigma^\mu d\sigma^\nu &\to \eta_{\mu\nu }d\sigma'^\mu d\sigma'^\nu =\Omega^2\eta_{\mu\nu }d\sigma^\mu d\sigma^\nu \ ,
\end{align}   
\end{subequations}
for some conformal factor $\Omega$ with a scaling transformation
\begin{subequations}
\begin{align}
    Y^M&\to \Omega^{1/\alpha'}Y^M\nonumber \ , \\
    x^i&\to \Omega^{\gamma'/\alpha'}x^i\nonumber \ , \\
    X^I&\to \Omega^{\beta'/\alpha'} X^I \ .
\end{align}
\end{subequations}
This is clearly a symmetry if $\Omega$ is constant as it is equivalent to the transformation in (\ref{eq: rigidC}). However, the full conformal group includes transformations where $d\Omega=\partial_\mu\Omega d\sigma^\mu\ne 0$: in these cases the above transformations will induce cross terms of the form $d\sigma^\mu dX^I$, $d\sigma^\mu dx^i$, and $d\sigma^\mu dY^M$  into the metric. If $\beta'=0$, however, then there will be no $d\sigma^\mu dX^I$ cross terms in ${\cal T}$, leaving only  $d\sigma^\mu dx^i$ and $d\sigma^\mu dY^M$ cross terms in ${\cal E}$. These, as we have discussed above, can be removed by local Galilean boosts, and so in this case the full conformal group of Minkowski spacetime is promoted to a symmetry of the NC geometry\footnote{Similarly, if $\beta'\ne 0 $ and $\alpha'=0$ then the Euclidean conformal group associated to the $X$-plane can be promoted to a symmetry of the whole NC geometry.}. This also occurs if there are no $X^I$ coordinates to scale. The Galilean boost symmetry therefore a powerful mechanism for symmetry-enhancement over what one would find in Lorentzian geometry. It is interesting to note that there is still an $AdS$ factor in the metric $g$, but it is split between ${\cal T}$ and $\cal E$, and in some cases the associated conformal symmetries of $AdS$ still persist in the dynamics.

Lastly, we comment that  String Theory and M-theory  posses other fields and these must also remain invariant under any symmetry of the metric (up to potential gauge transformations). In particular, String Theory  has a dilaton field that takes the form
\begin{align}
e^{\Phi} = c^{3-q}e^{\phi}	\ ,
\end{align}
in a D$q$NC limit. This will generically break the conformal symmetries discussed above unless it is constant following the rescaling $c\to\Lambda c$ used above, as is familiar from standard $AdS$ backgrounds in String Theory.

\subsection{Newton-Cartan Geometries from Intersecting Brane Metrics}

Let us start with the intersecting D-brane solution ($t=\sigma^0$):
\begin{subequations} \label{eq: intersecting d-brane solution} 
\begin{align} \nonumber
    g & = H^{-\frac12}_pH^{-\frac12}_q	\eta_{\mu\nu}d\sigma^\mu d\sigma^\nu +H^{-\frac12}_p  H^{\frac12}_qdx\cdot d x \\
    & \qquad + H^{\frac12}_pH^{-\frac12}_q d X\cdot dX+ H^{\frac12}_pH^{\frac12}_q dY\cdot d Y \ , \\
    C_{p+1} & = H_p^{-1} d \sigma^0\wedge \ldots  \ldots dX^p \ , \\
    C_{q+1} & = H_q^{-1} d\sigma^0\wedge  \ldots \ldots \wedge dx^qr \ , \\
    e^\Phi & = H_p^{\frac{3-p}{4}}H_q^{\frac{3-q}{4}} \ .
\end{align}
\end{subequations}
To solve the supergravity solutions one brane must be smeared along the directions of the other. In our case we want to consider a D$q$NC limit corresponding to completely smearing the D$q$-branes. In particular we let $H_q=c^{-4}$ so that far away from the D$p$-branes, where  $H_p\sim 1$, the metric becomes flat space but in a D$q$NC form:
\begin{align}
	g \to c^2\left[ \eta_{\mu\nu}d\sigma^\mu d\sigma^\nu + dX\cdot d X\right] + c^{-2} \left[   d x\cdot dx+  dY\cdot d Y\right] \ .
\end{align}
The equation for $H_p$ is
\begin{align} 
    H_q \, \partial_X\cdot\partial_X H_p + \partial_Y\cdot \partial_Y H_p&=0 \ .
\end{align}
In general, when $H_q$ is not constant, this equation is solved by taking $H_p$ to be smeared along the $X$-directions:
\begin{align}\label{Hsmeared}
	H_p = 1 + \frac{R^{n-2}}{ |Y|^{n-2}} \ ,
\end{align}
where $n\le 9-p$ is the  number of dimensions transverse to both branes, parameterized by the coordinates $Y^M$. This is the solution we will take in this paper. As a result, in the near-horizon limit we find
\begin{subequations}
\begin{align} \nonumber
    g &=	c^2\left[\left(\frac{R}{|Y|}\right)^{-\frac{n-2}{2}}\eta_{\mu\nu }d\sigma^\mu d\sigma^\nu+\left(\frac{R}{|Y|}\right)^{\frac{n-2}{2}}dX\cdot dX \right] \\
    & \qquad + c^{-2} \left[\left(\frac{R}{|Y|}\right)^{-\frac{n-2}{2}}dx\cdot dx    + \left(\frac{R}{|Y|}\right)^{\frac{n-2}{2}}dY\cdot dY \right] \ , \\
    e^{\phi} & = c^{q-3}\left(\frac{R}{|Y|}\right)^{\frac{(3-p)(n-2)}{4}}\ .   
\end{align}
\end{subequations}
which is indeed of the form discussed above in (\ref{eq: NCg}). We will also consider intersection brane solutions in M-theory where a similar metrics  appear. The details will be presented below.

\subsubsection{To Smear Or Not To Smear}

In a previous paper \cite{Lambert:2024uue,Lambert:2024yjk,} we took advantage of the fact $H_q=c^{-4}$ is constant to give an alternative, localized, solution for the harmonic function that appears by rescaling the initial supergravity solution,
\begin{align}\label{Hpis}
    H_p = 1 + \frac{R^{7-p}}{(|X|^2+ c^{-4}|Y|^2)^{\frac{7-p}{2}}} \ .
\end{align}
In this case, starting from the intersecting D-brane metric, we find the near horizon geometry 
\begin{subequations}
\begin{align} \nonumber
    g & = c^2\left[\left(\frac{|X|}{R}\right)^{\frac{7-p}{2}}\eta_{\mu\nu}dx^\mu dx^\nu +\left(\frac{R}{|X|}\right)^{\frac{7-p}{2}} dX\cdot dX \right] \\
    & \qquad +	c^{-2}\left[\left(\frac{|X|}{R}\right)^{\frac{7-p}{2}}dx\cdot dx + \left(\frac{R}{|X|}\right)^{\frac{7-p}{2}}dY\cdot dY \right] \ , \\
    e^\phi &=  c^{q-3}\left(\frac{R}{|X|}\right)^{\frac{(7-p)(3-p)}{4}} \ ,
\end{align}
\end{subequations}
where we have dropped subleading terms in ${\cal T}$ and ${\cal E}$, {\it i.e.} further terms that are suppressed by factors of $c^{-4}$.
In this case we can use the NCW transformation \eqref{eq: ncw general transformation} to make the large dimensions take the form of $AdS\times S$,
\begin{subequations}
\begin{align}\nonumber
    g' &   = c^2\left[\left(\frac{|X|}{R}\right)^{5-p}\eta_{\mu\nu}dx^\mu dx^\nu +\left(\frac{R}{|X|}\right)^{2} dX\cdot dX \right]
    \\
    & \qquad +	c^{-2}\left[\left(\frac{|X|}{R}\right)^{ 2}dx\cdot dx + \left(\frac{R}{|X|}\right)^{5-p}dY\cdot dY \right] \ ,
    \\
    e^{\phi'} &=  c^{q-3}\left(\frac{R}{|X|}\right)^{\frac{ (3-p)(10-p-q)}{4}} \ .
\end{align}
\end{subequations}

By the same arguments we discussed in section \ref{sect: General} such metrics also admit the full conformal group of the Minkowski space  (which in general is broken by the dilaton). In particular, while the rigid scale transformations of $AdS$ trivially extend to the whole metric for the special conformal transformations one needs to use local Galilean transformations to extend the symmetries to the whole metric due to the presence of warp factors.  

The difference between the two choices of harmonic function can be thought of as arising from how we scale the number of D$p$-branes as a function of $c$. To see this we start in an asymptotically Minkowskian spacetime with no NC deformation. Here we can consider two types of solutions if $H_q=1$:
\begin{subequations}\label{2Hs}
\begin{align}
    H^{smeared}_p & = 1 + \frac{R^{n-2}_s}{|Y|^{n-2}} \ , \\
    H^{localised}_p &= 1 + \frac{R^{7-p}_l}{(|X|^2+  |Y|^2)^{\frac{7-p}{2}}} \ .
\end{align}
\end{subequations}
As is well known the variable $R_l$ is related to the number of D$p$-branes and string coupling by $R^{7-p}_l\sim g_s N_p$, where we will ignore unimportant numerical constants. On the other hand, in the smeared case one has $ R^{n-2}_s\sim g_s N_p/V_s$ where $N_p/V_s$ is the density of D$p$-branes per unit of smeared volume. 

To implement the D$q$NC limit from the solutions in (\ref{2Hs}) we rescale $X\to cX$ and $Y\to c^{-1}Y$, but also
take
\begin{align}
 R_l= cR \qquad R_s = c^{-1}R  \qquad g_s  = c^{q-3}g\ ,
\end{align}
with $R$ and $g$ fixed.  Thus in the localised case we have
\begin{align}
N_p\sim g_s^{-1}R_l^{7-p} = c^{10-p-q}g^{-1}R^{7-p}   \ , 
\end{align}
and hence we are sending the number of D$p$-branes to infinity (since $p+q=4$ in all our examples). 
On the other hand we can consider the smeared case. Here  the volume $V_s$ is also rescaled to $V_s=c^{9-p-n}V$, with $V$ held fixed. We therefore have
\begin{align}
N_p \sim g_s^{-1}R_{s}^{n-2} V_s =c^{3-q}c^{2-n} c^{9-p-n} g^{-1}R^{n-2}V    = c^{14-q-2n-p} g^{-1}R^{n-2}V \ .
\end{align}
In all cases we discuss below we have $p+q=4$ and $n=5$ so $N_p\sim  g^{-1}R^{n-2}V$, so the smeared case does not require a rescaling of the number of D$p$-branes.

In M-theory there is no dilaton and the argument is a little more simple. For localised solutions the number of branes is given by $N_p\sim c^{8-p}R^{8-p}$, which diverges as we take the $c\to\infty$ limit. Let us now consider the smeared cases. In an M2NC limit one has 
\begin{align}
    X\to cX\qquad Y \to c^{-\frac12}Y  \ ,
\end{align}
so $R_l = cR$, $R_s = c^{-\frac12}R$ and $V_s = c^{10-p-n}V$, resulting in
\begin{align}
    N_p \sim R_s^{n-2}V_s = c^{11-p-3n/2}R^{n-2}V \ ,
\end{align}
for smeared branes. This is independent of $c$ for $p=5$ and $n=4$. In an M5NC limit we have
\begin{align}
    X\to cX\qquad Y \to c^{-2}Y \ ,
\end{align}
so $R_l = cR$, $R_s = c^{-2}R$ and $V_s = c^{10-p-n}V$; the same argument now gives
\begin{align}
    N_p \sim R_s^{n-2}V_s = c^{14-p-3n}R^{n-2}V \ ,
\end{align}
for smeared branes, and again for $p=2$ and $n=4$ we find this is independent of $c$. In the remainder of this paper we will only consider smeared solutions. It would be interesting to better understand the role of the localised case, which involves an additional rescaling of the rank of the gauge theory, but that is beyond the scope of this paper.

\section{The D0-D4 System}\label{sect: d0-d4}

The first system that we will consider is the bound state of D0-branes and D4-branes formed by the brane configuration
\begin{align}
\begin{array}{r cccccc}
	D0:& 0 & & & &  \\
    D4:& 0 & 1 &  2& 3& 4 
\end{array} \ .
\end{align} 
We shall consider taking the joint D0NC-D4NC limit of the set-up from the perspective of both the field theory and the dual gravitational picture for both orders of limits. The physical interpretation of this joint limit is similar to that of a regular D$p$NC limit, where we isolate the near-BPS dynamics of the brane, except that the BPS bound in question is that of the bound-state. From the string theory embedding it is clear that the two limits should commute; in the field theory we shall find that this manifests itself as the equality of the D0NC limit of a five-dimensional theory and the D4NC limits of a one-dimensional theory. This is essentially a consequence of the well-known fact that the ADHM correspondence \cite{Atiyah:1978ri} is given a string-theoretic embedding by the D0-D4 system \cite{Witten:1994tz, Douglas:1996uz}. As this system is well studied already we shall only consider the Bosonic sector here, though there is no barrier to also including Fermions.

\subsection{Field Theory Decoupling Limits}

\subsubsection{The D0NC Limit of Five-Dimensional \texorpdfstring{$\cal{N}=2$}{N=2} Super Yang-Mills} \label{sect: D0NC-D4}


Let us start by considering the D0NC limit of five-dimensional $\cal{N}=2$ SYM. As this arises by taking the longitudinal D4NC limit of the worldvolume theory on D4-branes the resulting theory can be thought of as a multi-critical decoupling limit \cite{Blair:2023noj,Blair:2024aqz}. The D0-brane supergravity solution of type IIA supergravity in the string frame is
\begin{subequations}
\begin{align}
    G &= - H_0^{-1/2} dt^2 + H_0^{1/2} \brac{dx^i dx^i + dY^M dY^M} \ , \\
    e^{\varphi} &= H_0^{3/4} \ , \\
    C_1 &= H_0^{-1} dt \ ,
\end{align}
\end{subequations}
where $H_0$ is harmonic in the transverse coordinates $(x^i, Y^M)$. We therefore implement the D0NC limit by taking
\begin{equation}
    H_0 = c^{-4} \ ,
\end{equation}
before taking the limit $c\to\infty$. The Bosonic action that we must consider is
\begin{align}
    \hat{S}_{\cal{N}=2} = \inv{2\hat{g}_{YM}^2} \tr \int d\hat{t} d^4\hat{x} \bigg( 
    -\inv{2} \hat{F}_{\mn} \hat{F}^{\mn} - \hat{D}_{\mu} \hat{Y}^M \hat{D}^{\mu} \hat{Y}^M + \inv{2} [\hat{Y}^M, \hat{Y}^N]^2 \bigg) \ .
\end{align}
The D0NC limit then corresponds to the scaling
\begin{subequations}
\begin{align}
    \hat{t} &= c t \ , \\
    \hat{x}^i &= c^{-1} x^i \ ,
\end{align}
\end{subequations}
of our coordinates, and
\begin{subequations}
\begin{align}
    \hat{Y}^M(\hat{t},\hat{x}) &= c^{-1} Y^M(t,x) \ ,
\end{align}
\end{subequations}
of the 5 scalars. The scaling of the gauge field's components is fixed by requiring that gauge-invariance is maintained. The scaling of the dilaton indicates that we must also rescale the Yang-Mills coupling to $\hat{g}_{YM}^2 = c^{-3} g_{YM}^2$. The field theory's action then becomes
\begin{equation}
    \hat{S}_{D4} = \inv{2g_{YM}^2} \tr \int dt d^4x \bigg( 
    - \frac{c^4}{2} F_{ij} F_{ij} + F_{0i} F_{0i} - D_i Y^M D_i Y^M
    \bigg) + O(c^{-4}) \ .
\end{equation}
The divergent part of this can be written as
\begin{equation}
    \tr\brac{F_{ij} F_{ij}} = \tr\brac{\inv{2}\brac{F_{ij} - *F_{ij}}^2 + \inv{2} \epsilon_{ijkl} F_{ij} F_{kl} } \ ,
\end{equation}
where the Hodge dual is taken with respect to the flat metric on the spatial coordinates. The second term is a total derivative, and is in fact cancelled after coupling in the D0NC flat connection
\begin{equation}
    C_1 = c^4 dt 
\end{equation}
 via the original D4-brane theory's Wess-Zumino term\footnote{It is straightforward to see that this term survives the D4NC limit for this choice of $C_1$.}
\begin{align} \nonumber
    \hat{S}_{WZ} &= \frac{T_4 \brac{2\pi \alpha'}^2}{2} \int C_1 \wedge \tr \brac{F \wedge F} \\
    &\equiv \frac{c^4}{8 g_{YM}^2} \tr \int dt d^4x \, \epsilon_{ijkl} F_{ij} F_{kl} \ ,
\end{align}
where we've used
\begin{equation}
    g_{YM}^2 = \frac{1}{(2\pi \alpha')^2 T_4} \ ,
\end{equation}
to relate the D4-brane tension and the Yang-Mills coupling. As the divergent term is a squared quantity, we can introduce a Hubbard-Stratonovich field $G_{ij}$ to rewrite it as
\begin{equation}
    \frac{c^4}{2} \tr \int dt d^4 x \brac{F_{ij} - *F_{ij}}^2 \Rightarrow \tr \int dt d^4x \brac{
    G_{ij}  \brac{F_{ij} - *F_{ij}} - \frac{c^{-4}}{2} G_{ij} G_{ij}
    } \ ,
\end{equation}
so taking the $c\to\infty$ limit gives the action
\begin{align} \label{eq: D0NC action}
    S_{D0NC} = \inv{2g_{YM}^2} \tr \int dt d^4x \bigg( 
     F_{0i} F_{0i} + G_{ij} \brac{F - * F}_{ij} - D_i Y^M D_i Y^M
    \bigg) \ .
\end{align}
This limit was first discussed in \cite{Lambert:2019nti}, with the reinterpretation as a D0NC limit giving a physical interpretation to the choice of one sign in the constraint over the other.

The theory is simple in this case: the Lagrange multiplier $G_{ij}$ imposes the instanton equations
\begin{equation} \label{eq: instanton equations}
    F = * F \ ,
\end{equation}
with the dynamics of the theory reducing to supersymmetric quantum mechanics on their moduli space. The spacetime symmetries of this action were discussed in \cite{Lambert:2019fne}, forming the Schr\"odinger group $Schr(4)$, and the $SO(5)$ R-symmetry of the relativistic theory survives the D0NC limit\footnote{For the purely Bosonic actions we will find that the $SO(5)$ R-symmetry rotations can take on arbitrary time-dependence. However, once the Fermionic sector is included this will not be true and so we will not discuss such enhancements any further in this paper.}. Of particular interest is the $SO(2,1)$ subgroup corresponding to one-dimensional conformal transformations. From the perspective of the quantum-mechanical theory this arises from the special K\"ahler structure of instanton moduli space. In particular, invariance under scale transformations requires a so-called  homothetic Killing vector $k_I$ that satisfies ${\cal L}_Kg_{IJ}=-2g_{IJ}$. To further enhance the symmetry to $SO(2,1)$ including a special conformal transformation means that $k_I=\partial_I  K$ for some potential $K$  (for a discussion of the special geometry see \cite{Vandoren:2000qr}). 

The non-dynamical scalar fields $Y^A$ obey the equations of motion
\begin{equation} \label{eq: Y spatial laplacian}
    D_i D_i Y^M = 0 \ ,
\end{equation}
with specified values at the asymptotic spatial boundary. Their role is to provide a potential for the quantum-mechanical theory; as \eqref{eq: Y spatial laplacian} is obeyed this can be computed using only boundary data. The potential can also be thought of as the norm of a triholomorphic Killing vector on the moduli space, as reviewed in appendix \ref{sect: appendix moduli space d0-d4}.

\subsubsection{The D4NC Limit of the Berkooz-Douglas Matrix Model}


Let us next discuss the system from the perspective of the D0-branes by reversing the order of the D0NC and D4NC limits. Unlike the previous case, the D4NC limit does not just isolate the low-energy degrees of freedom on the D0-branes described by the BFSS matrix model; one must also include the low-energy degrees of freedom stemming from strings stretching between both types of branes, leading to the Berkooz-Douglas matrix model \cite{Berkooz:1996is}. These additional modes are not required in other limits as the intersection is of lower dimension on the brane and such additional open stretched strings only affect possible boundary conditions. However, for the D0-D4 system the intersection is the entire D0-brane and these modes are important for the system's dynamics. For now we will consider the case for which the D4-branes are all coincident. This has the one-dimensional $U(k)$ gauge theory action
\begin{align} \nonumber
    \hat{S}_{D0} = \inv{2\hat{g}_{YM}^2} \tr \int d\hat{t} \, \bigg( & \hat{D}_t \hX^A \hat{D}_t \hX^A + 2 \hat{D}_t \hcalz_a^{\alpha} \hat{D}_t \hcalz^{\dag \, \alpha a} + \hat{D}_t \hY^M \hat{D}_t \hY^M 
    \\ \nonumber
    &- \inv{2} \brac{\hcalz^{\alpha}_a \sigma^i_{\beta\alpha} \hcalz^{\dag\, \beta  a} - i \bar{\eta}^i_{AB} [\hX^A, \hX^B] }^2 + [\hX^A, \hY^M]^2
    \\ \label{eq: D0 brane action}
    & - 2  \hY^M\hcalz^{\alpha}_a  \hcalz^{\dag \, \alpha a}  \hY^M + \inv{2} [\hY^M, \hY^N]^2 \bigg) \ ,
\end{align}
where $\hcalz$ are scalar fields in the fundamental representations of $U(k)$ and $SU(2)$ (indexed by $\alpha,\beta$), and the antifundamental of $SU(N)$ (indexed by $a$). Note that in this section we label the spatial D4-brane directions with an index $A=1,\ldots,4$, with $i=1,2,3$ an auxiliary index. The D4NC limit of a flat background is given by
\begin{subequations}
\begin{align}
    G &= c^2 \brac{- dt^2 + dX^A dX^A} + c^{-2} dY^M dY^M \ , \\
    e^{\varphi} &= c \ , \\
    C_5 &= c^4 dt \wedge dX^1 \wedge \ldots \wedge dX^5 \ ,
\end{align}
\end{subequations}
and so the limit of the field theory corresponds to taking the coordinate along the brane to scale as
\begin{equation}
    \hat{t} = c t \ ,
\end{equation}
while taking the fields to have the scaling
\begin{subequations}
\begin{align}
    \hat{X}^A(\hat{t}) &= c X^A(t) \ , \\
    \hcalz_a^{\alpha}(\hat{t}) &= c \calz_a^{\alpha}(t) \ , \\
    \hY^M(\hat{t}) &= c^{-1} Y^M(t) \ , \\
    \hat{A}_0(\hat{t}) &= c^{-1} A_0(t) \ ,
\end{align}
\end{subequations}
and the coupling to scale as
\begin{equation}
    \hat{g}_{YM}^2 = c g_{YM}^2 \ .
\end{equation}
The contribution of the divergent 5-form spacetime form-field is trivial, as
\begin{equation}
    \epsilon_{ABCD} \tr \brac{[X^A,X^B][X^C,X^D]} = 0 \ .
\end{equation}
The potential in the middle line of \eqref{eq: D0 brane action} is proportional to $c^4$ after the rescaling and must therefore be regularised using a Hubbard-Stratonovich field $G^i$. Hence, taking the $c\to \infty$ limit yields
\begin{align} \nonumber
    S_{D4NC} = \inv{2g_{YM}^2} \tr \int dt \, \bigg( &
    \partial_t X^A \partial_t X^A + 2 \partial_t \calz_a^{\alpha} \partial_t \calz^{\dag \, \alpha a} \\ \nonumber
    & + G^i \brac{\calz_a^{\alpha} \sigma^i_{\beta\alpha} \calz^{\dag \, \beta a} - i \bar{\eta}^i_{AB} [X^A, X^B]} \\ \label{eq: D4NC action}
    &+ [X^A, Y^M]^2 - 2  Y^M\calz^{\alpha}_a  \calz^{\dag \, \alpha a}  Y^M \bigg) \ .
\end{align}
The constraints
\begin{equation} \label{eq: D4NC constraint}
    \calz_a^{\alpha} \sigma^i_{\alpha\beta} \calz^{\dag \, \beta a} - i \bar{\eta}^i_{AB} [X^A, X^B] = 0
\end{equation}
imposed by $G^i$ are nothing but the ADHM equations, and the theory reduces to quantum mechanics on their moduli space. This is related to instanton moduli space by a hyper-K\"ahler isometry \cite{Maciocia:1991ph}, meaning the kinetic terms of the two limit orders are equal. A necessary consequence of this is that the symmetries of the two actions must match, as can be easily checked using a slightly adapted form of the analysis in \cite{Aharony:1997an}.

A key difference between this action and \eqref{eq: D0NC action} is that if we attempt to impose the equation of motion for the transverse scalar fields $Y^M$ we find that it vanishes, and so the final two terms in \eqref{eq: D4NC action} evaluate to zero. We physically interpret the vanishing of $Y^M$ as telling us that the D0-branes and D4-branes coincide. However, this seemingly leaving us with no potential. The resolution is that the modes coming from the D0-D4 strings also couple to the $\frak{su}(N)$-valued fields $\Tilde{Y}^M$ denoting the positions of the D4-branes, which appear as matrix-valued parameters from the perspective of the D0-brane worldvolume theory. After introducing these and imposing the equation of motion for $Y^M$ the action becomes
\begin{align} \nonumber
    S_{D4NC} = \inv{2g_{YM}^2} \tr \int dt \,  \bigg( &  
    D_t X^A D_t X^A + 2 D_t \calz_a^{\alpha} D_t \calz^{\dag \, \alpha a} \\ \nonumber
    & + G^i \brac{\calz_a^{\alpha} \sigma^i_{\beta\alpha} \calz^{\dag \, \beta a} - i \bar{\eta}^i_{AB} [X^A, X^B]}  \\ \label{eq: D4NC action 2} 
    &- 2  \calz^{\alpha}_a \tensor{(\Tilde{Y}^M)}{^a_c} \tensor{(\Tilde{Y}^M)}{^c_b} \calz^{\dag \, \alpha b} \bigg) \ .
\end{align}
The potential induced by non-zero values of $\Tilde{Y}^M$ is identical to that found in the previous case, which we show in appendix \ref{sect: appendix moduli space d0-d4}. In both cases we find the exact same dynamical system, and hence the order of limits are completely equivalent.


\subsection{The Dual Gravitational Picture}

\subsubsection{The D0NC Limit of the D4-Brane}


It is somewhat noteworthy that taking the D0NC limit of an initially non-conformal theory reduces it to one containing the one-dimensional conformal group. We would like to understand the dual gravitational picture of this process, given by the D0NC limit of type IIA supergravity. A full description of the limit is given in appendix \ref{sect: iia d0nc action}; for our purposes we will only need that the limit splits the relativistic fields into the variables
\begin{subequations} \label{eq: D0NC decomp}
\begin{align}
    \hat{G}_{\mn} &= - c^2 \tau_{\mu} \tau_{\nu} + c^{-2} \cE_{\mn} \ , \\
    \hat{G}^{\mn} &= c^2 \cE^{\mn} - c^{-2} \tau^{\mu} \tau^{\nu} \ , \\
    e^{\hat{\varphi}} &= c^{-3} e^{\varphi} \ , \\
    \hat{C}_1 &= c^4 e^{-\varphi} \tau + C_1 \ , \\
    \hat{C}_3 &= C_3 \ , \\
    \hat{B}_2 &= B_2 \ ,
\end{align}
\end{subequations}
resulting in a 0-brane Newton-Cartan geometry with local Galilean boost and Newton-Cartan-Weyl symmetries. Using the harmonic function rules for $p$-brane Newton-Cartan limits discussed in section \ref{sect: General}, the D4-D0NC solution of the relativistic theory is
\begin{subequations}
\begin{align}
        \hat{G} &= - c^2 H_4^{-1/2} dt^2 + c^{-2} \brac{ H_4^{-1/2} dx^i dx^i +  H_4^{1/2} dY^M dY^M } \ , \\
        e^{\hat{\varphi}} &= c^{-3} H_4^{-1/4} \ , \\
        \hat{C}_1 &= c^4 dt \ , \\
        \hat{C}_5 &= \brac{H_4^{-1} - 1} dt \wedge dx^1 \wedge \ldots \wedge dx^4 \ , \\
        \hat{B}_2 &= 0 \ ,
\end{align}
\end{subequations}
where $H_4$ is the harmonic function
\begin{align}
    H_4 = 1 + \frac{R^3}{|Y|^3} \ .
\end{align}
This gives us a geometry of the form \eqref{eq: D0NC decomp}, with
\begin{subequations}
\begin{align}
    \tau &= H_4^{-1/4} dt \ , \\
    \cE &= H_4^{1/2} \partial_i \otimes \partial_i + H_4^{-1/2} \partial_M \otimes \partial_M \ , \\
    e^{\varphi} &= H_4^{-1/4} \ , \\
    C_1 &= 0 \ , \\
    C_5 &= \brac{H_4^{-1} - 1} dt \wedge dx^1 \wedge \ldots \wedge dx^4 \ , \\
    B_2 &= 0 \ .
\end{align}
\end{subequations}
It will be convenient to work in a gauge where the value of the dilaton is fixed to zero, which is achieved using the Newton-Cartan-Weyl transformation \eqref{eq: NCW transformation} with parameter
\begin{equation}
    e^{\omega} = H_4^{-1/12} \ ;
\end{equation}
in this gauge we have
\begin{subequations}
\begin{align}
    \tau &= H_4^{-1/3} dt \ , \\
    \cE &= H_4^{1/3} \partial_i \otimes \partial_i + H_4^{-2/3} \partial_M \otimes \partial_M \ .
\end{align}
\end{subequations}
Finally, taking the near-horizon limit leaves us with 
\begin{subequations} \label{eq: D4-D0NC gravity solution}
\begin{align}
    \tau &= \frac{|Y|}{R} dt \ , \\
    \cE &= \frac{R}{|Y|} \partial_i \partial_i + \frac{|Y|^2}{R^2} \partial_M \partial_M \ , \\
    C_1 &= 0 \ , \\
    C_5 &= \frac{|Y|^3}{R^3} dt \wedge dx^1 \wedge \ldots \wedge dx^4 \ , \\
    B_2 &= 0 \ .
\end{align}
\end{subequations}
The geometry is the product of a Newton-Cartan version of the Lifshitz-$AdS_6$ spacetime \cite{Kachru:2008yh} and an $S^4$ factor. The D0NC limit of IIA can also be constructed as the null reduction of eleven-dimensional supergravity using the field decomposition (here $\odot$ denotes the symmeterized tensor product)
\begin{subequations} \label{eq: 11d sugra null reduction field def}
\begin{align}
    \hat{g} &= - 2  \tau \odot \brac{dx^+ - C_1} + \cE_{\mn} dx^{\mu} \otimes dx^{\nu} \ , \\ \label{eq: 11d null reduction form field ansatz}
    \hat{C}_3 &= C_3 - dx^+ \wedge B_2 \ ,
\end{align}
\end{subequations}
as discussed in appendix \ref{sect: iia d0nc action}. Similarly, the D4-D0NC near-horizon solution arises as the null reduction of the M5-brane's near-horizon geometry. The relativistic eleven-dimensional solution we're interested in is
\begin{subequations}
\begin{align}
    \hat{G} &= \frac{r}{R} \brac{- 2 dx^+ dx^- + dx^i  dx^i} + \frac{R^2}{|Y|^2} dY^2 + R^2 g_{S^4} \ , \\
    \hat{C}_6 &= \frac{|Y|^3}{R^3} dx^+ \wedge dx^- \wedge dx^1 \wedge \ldots \wedge dx^4 \ ;
\end{align}
\end{subequations}
taking the null direction $x^+$ to be compact and using the ansatz \eqref{eq: 11d sugra null reduction field def}\footnote{Here we use the dual of the ansatz \eqref{eq: 11d null reduction form field ansatz} for the three-form field's decomposition.} to identify the fields of the Newton-Cartan theory exactly recovers the solution \eqref{eq: D4-D0NC gravity solution}, as expected.

Let us analyse the symmetries of the solution. Taking $f(t)$ to be an arbitrary infinitesimal function of $t$, we see that fields are invariant under the coordinate transformation
\begin{subequations}
\begin{align}
    \hat{t} &= t + f \ , \\
    \hat{x}^i &= \brac{1 + \inv{2} \dot{f}} x^i \ , \\
    \hat{Y}^M &= \brac{1 - \dot{f}} Y^M \ .
\end{align}
\end{subequations}
In order for this to be a symmetry of the inverse NC metric fields, for which we take the representatives
\begin{subequations}
\begin{align}
    \tau &= - \frac{R}{|Y|} \partial_t \ , \\
    \cE &= \frac{|Y|}{R} dx^i dx^i + \frac{R^2}{|Y|^2} dY^M dY^M \ ,
\end{align}
\end{subequations}
we must also include the local Galilean boost
\begin{equation}
    \lambda = \ddot{f} \brac{ \frac{R^3}{|Y|^3} Y^M dY^M - \inv{2} x^i dx^i } \ .
\end{equation}
However, from \eqref{eq: local g-boost D0NC} this acts non-trivially on $C_1$; the transformation is therefore only a symmetry when $\lambda$ is a gauge transformation, which restricts $f(t)$ to
\begin{equation}
    f(t) = a + b t + c t^2 \ .
\end{equation}
Happily, we recover the $SO(2,1)$ symmetry group found in the field theory.

The other symmetries of the solution \eqref{eq: D4-D0NC gravity solution} are those that don't transform $t$, namely
\begin{subequations}
\begin{align}
    \hat{x}^i &= x^i + \tensor{r}{^i_j}(t) x^j + k^i(t) \ , \\
    \hat{Y}^M &= Y^M + \tensor{\cal{R}}{^M_N}(t) Y^M \ ,
\end{align}
\end{subequations}
where both $r$ and $\cal{R}$ are antisymmetric in their respective indices. As above, in order for these to be a symmetry of the inverse NC metrics and $C_1$ we must introduce the local Galilean boost
\begin{equation}
    \lambda = \frac{R^2}{|Y|^2} \tensor{\dot{\cal{R}}}{^M_N} Y^N d Y^M - \brac{\tensor{\dot{r}}{^i_j} x^j + \dot{k}^i} dx^i \ ,
\end{equation}
which must be closed. This is the case when both $r$ and $\cal{R}$ are constant and $k^i$ is linear in $t$, giving us the $ISO(4)$ and boost spatial symmetries as well as the $SO(5)$ R-symmetry found in the field theory. The full symmetry algebra is therefore
\begin{equation}
    \frak{g} = \frak{schr}(4) \oplus \frak{so}(5) \ ,
\end{equation}
which up to the global gauge-group factor exactly matches the field theory's symmetries. As the supergravity solution is only valid in the large $N$ limit it is unclear whether we should expect to see this part of the symmetry from the gravity theory.

In the relativistic theory, there were two factors that forbade the existence of a scaling symmetry in the near-horizon limit. The first is the structure of the $Y^M$-components of the metric. The metric's components are only functions of the radius $|Y|$, so any scaling must act non-trivially on $Y^M$. However, the $Y^M$-components are not invariant under a scaling of $Y^M$, and as such we cannot have a scaling symmetry in the relativistic theory. The second problem is the presence of a non-trivial dilaton which depends only on $|Y|$; since any scaling must scale the radius it will necessarily change the dilaton and is therefore not a symmetry. We see that both problems are solved in the same way by taking the D0NC limit; the existence of the local dilatation symmetry \eqref{eq: NCW transformation} allows us to trivialise the dilaton, with the transformation bringing the $Y^M$-components of $\cE$ into a form where they are invariant under scaling transformations of $Y^M$.

\subsubsection{The D4NC Limit via Eleven-Dimensional Supergravity}

In order to source a supergravity background for which we can take the D4NC limit we must consider a configuration with a large number of D0-branes. This means that the system is most naturally considered in M-theory\footnote{Indeed, taking a near horizon limit of the D0-D4 type IIA solution as we did above does not seem to lead to a meaningful gravitational dual in this case.}. We work with the compactified pp-wave spacetime
\begin{equation}
    ds^2 = - 2dx^+ dx^- + H_{pp} (dx^+)^2 + dx^i dx^i + dY^M dY^M \ ,
\end{equation}
where $x^+$ is periodic with period $2\pi R$ and $H_0$ is harmonic in the $Y^M$ directions, which upon using the decomposition
\begin{equation} \label{eq: 11d reduction}
    g_{11} = e^{4\phi/3} \brac{dx^+ - C_1}^2 + e^{-2\phi/3} g_{10} 
\end{equation}
of the eleven-dimensional metric into the ten-dimensional variables recovers the smeared D0-brane solution of interest with $H_0 = H_{pp}$. Taking the near-horizon limit of the D0-branes reduces $H_{pp}$ to
\begin{equation}
    H_{pp} = \frac{R^3}{|Y|^3} \ .
\end{equation}
The M-theory analogue of the D0-D4 intersecting brane solution is a pp-wave propagating along an M5-brane, with the metric and six-form field
\begin{subequations}
\begin{align}
    ds^2 &= H_5^{-1/3} \brac{-2 dx^+ dx^- + H_{pp} (dx^+)^2 + dx^i dx^i} + H_5^{2/3} dY^M dY^M \ , \\
    C_6 &= - H_5^{-1} dx^+ \wedge dx^- \wedge dx^2 \wedge \ldots \wedge dx^5 \ .
\end{align}
\end{subequations}
We can then implement the M5NC limit by setting $H_5 = c^{-6}$ and taking the limit $c\to\infty$; using \eqref{eq: 11d reduction} it is straightforward to see that this corresponds to the D4NC limit of the ten-dimensional reduction. The Newton-Cartan solution we find after the limit is
\begin{equation} \label{eq: d0d-d4 m5nc gravity}
    ds^2 = c^2 \brac{ - 2 dx^+ \brac{dx^- - \frac{R^3}{2|Y|^3} dx^+ } + dx^i dx^i } + c^{-4} dY^M dY^M \ .
\end{equation}
The $(dx)^2$ term is subleading in the large-$|Y|$ limit, and we asymptotically find a solution of the form
\begin{equation} \label{eq: d0-d4 m5nc gravity asymptotic}
    ds^2 \to c^2 \brac{ - 2dx^+ dx^- + dx^i dx^i} + c^{-4} dY^M dY^M \ .
\end{equation}
Using the proposed Galilean boost and Newton-Cartan-Weyl scaling symmetries of the M5NC limit we see that the symmetries of \eqref{eq: d0-d4 m5nc gravity asymptotic}\footnote{Which are therefore the asymptotic symmetries of \eqref{eq: d0d-d4 m5nc gravity}.} at the $|Y|\to\infty$ boundary are the conformal symmetries of $\cal{T} = \eta_{1,5}$ and the $SO(5)$ R-symmetry rotations of $Y^M$. As $x^+$ is compact the $SO(2,6)$ conformal group of six-dimensional Minkowski spacetime are broken to $Schr(4)\times U(1)$. The $Schr(4)\times SO(5)$ factors exactly match the symmetry of the D0NC gravitational limit considered above, while the $U(1)$ factor is associated to the field theory's conserved instanton-number current and is not apparent from the ten-dimensional approach.

\section{The D1-D3 System} \label{sect: d1-d3}

The next system of interest is the bound state of D1-branes and D3-branes, which arises from the brane configuration
\begin{align}
\begin{array}{r cccccc}
	D1:& 0 & & & & 4  \\
    D3:& 0 & 1 &  2& 3&  
\end{array} \ .
\end{align} 
Though this is related to the D0-D4 system through T-duality the analysis will be somewhat different and therefore worth analysing in its own right. The D1-branes in the bound state are of finite extent, and are stretched along the fourth spatial direction between D3-branes. We will find that the joint D1NC-D3NC limit of the string theory configurations leads to an equality between the D1NC limit of a four-dimensional theory and the D3NC limit of a one-dimensional theory. As for the D0-D4 system in section \ref{sect: d0-d4}, this arises as a consequence of the string theory embedding of the Nahm equation/BPS monopole correspondence \cite{Diaconescu:1996rk}.

\subsection{Field Theory Decoupling Limits}

\subsubsection{The D1NC Limit of Four-Dimensional \texorpdfstring{$\cal{N}=4$}{N=4} Super Yang-Mills}


We begin by taking the D3NC limit of the system first, leading to a description of the bound state within four-dimensional $\cal{N}=4$ SYM. The D1NC limit of this theory was studied in detail in \cite{Lambert:2024yjk} and so our review here will be brief.

The Bosonic action for $\cal{N}=4$ SYM is
\begin{align} \nonumber
    \hat{S}_{\cal{N}=2} = \inv{2\hat{g}_{YM}^2} \tr \int d\hat{t} d^3\hat{x} \bigg( &
    -\inv{2} \hat{F}_{\mn} \hat{F}^{\mn} - \hat{D}_{\mu} \hat{X} \hat{D}^{\mu} \hat{X} - \hat{D}_{\mu} \hat{Y}^M \hat{D}^{\mu} \hat{Y}^M \\
    &+ [\hat{X}, \hat{Y}^M]^2 + \inv{2} [\hat{Y}^M, \hat{Y}^N]^2 \bigg) \ .
\end{align}
As the D1NC of a flat background takes the form
\begin{subequations}
\begin{align}
    G &= c^2 \brac{- dt^2 + dX dX} + c^{-2}\brac{dx^i dx^i + dY^M dY^M }\ , \\
    e^{\varphi} &= c^{-1} \ , \\
    C_2 &= c^4 dt \wedge dX \ ,
\end{align}
\end{subequations}
we should take the coordinates and fields to scale as
\begin{subequations}
\begin{align}
    \hat{t} &= c t \ , \\
    \hat{x}^i &= c^{-1} x^i \ ,
\end{align}
\end{subequations}
and
\begin{subequations}
\begin{align}
    \hat{X}(\hat{t}, \hat{x}) &= c X(t,x) \ , \\
    \hat{Y}^M(\hat{t}, \hat{x}) &= c^{-1} Y^M(t,x) \ , \\
    \hat{A}_0(\hat{t}, \hat{x}) &= c^{-1} A_0(t,x) \ , \\
    \hat{A}_i(\hat{t}, \hat{x}) &= c A_i(t,x) \ .
\end{align}
\end{subequations}
We must also rescale the Yang-Mills coupling to
\begin{equation}
    \hat{g}_{YM}^2 = c^{-1} g_{YM} \ ,
\end{equation}
and include the Wess-Zumino term associated to the divergent two-form field,
\begin{equation}
    S_{WZ} = \frac{c^4}{2 g_{YM}^2} \tr \int dt d^3 x \, \epsilon_{ijk} D_i X F_{jk} \ .
\end{equation}

Taking the $c\to\infty$ limit then takes us to the action
\begin{align} \nonumber
    S_{D1NC} = \inv{2 g_{YM}^2} \tr \int dt d^3x \, \Big(&
    F_{0i} F_{0i} + D_t X D_t X + G_{ij} \brac{F_{ij} - \epsilon_{ijk} D_k X} \\ \label{eq: D1NC action}
    &- D_i Y^M D_i Y^M + [X,Y^M]^2 \Big) \ .
\end{align}
By imposing the equation of motion for $G_{ij}$ we see that the spatial dependence of the dynamic fields is constrained to obey the BPS monopole equation
\begin{equation} \label{eq: monopole equations}
    F_{ij} = \epsilon_{ijk} D_k X  \ ,
\end{equation}
with the theory then reducing to quantum mechanics on the moduli space of solutions. The non-dynamical transverse scalars $Y^M$ satisfy the equation of motion
\begin{equation}
    D_i D_i Y^M - [X,[X,Y^M]] = 0 .
\end{equation}
As in the instanton theory their role is to source a potential, which can again be interpreted as the norm of a tri-holomorphic Killing vector on the moduli space of \eqref{eq: monopole equations}: we review this in appendix \ref{sect: appendix moduli space d1-d3}.

The action is invariant under an $SO(2,1)$ subgroup of the original $SO(2,4)$ conformal symmetry, but as any non-trivial solution of \eqref{eq: monopole equations} requires an asymptotic value for $X$ (which has a non-zero scaling dimension) the dilatation and special conformal symmetries are broken. Indeed, the associated moduli space metric does not admit a homothetic Killing vector that could generate a scale symmetry. After integrating out $G_{ij}$\footnote{If we take the action \eqref{eq: D1NC action} at face value we find that spatial translations can be given arbitrary time-dependence; however, we should interpret this as a gauge symmetry that is used to fix the extra degrees of freedom associated to the Lagrange multiplier $G_{ij}$.} and working directly on the constraint surface, we find the remaining spacetime symmetries are spatial translations, rotations, and Galilean boosts. The classical spacetime symmetries therefore form the Galilean group, which one would assume is extended to the Bargmann algebra upon quantization through the inclusion of the topological monopole charge as a central extension. We also find an $SO(5)$ R-symmetry rotating the transverse scalars, though this is spontaneously broken when $Y^M$ is given a non-zero asymptotic value.

\subsubsection{The D3NC Limit of Two-Dimensional \texorpdfstring{$\cal{N}=(8,8)$}{N=(8,8)} Super Yang-Mills}


The reciprocal picture we get by swapping the order of limits leads us to take the D3NC limit of two-dimensional $\cal{N}=(8,8)$ SYM, which we obtain by taking the D1NC limit of the D1-brane's worldvolume theory. The Bosonic sector of this theory has the action
\begin{align} \nonumber
    \hat{S}_{D1} = \inv{2g_{YM}^2} \tr \int d\hat{t} d \hat{x} \bigg(&
    \hat{F}_{01}^2 - \hat{D}_{\mu} \hat{X}^i \hat{D}^{\mu} \hat{X}^i  - \hat{D}_{\mu} \hat{X}^i \hat{D}^{\mu} \hat{X}^i +  \inv{2} [\hat{X}^i, \hat{X}^j]^2 \\ \label{eq: 2d MSYM action}
    &+ [\hat{X}^i, \hat{Y}^M]^2 + \inv{2} [\hat{Y}^M, \hat{Y}^N]^2
    \bigg) \ .
\end{align}
The D3NC limit of a flat background is
\begin{subequations}
\begin{align}
    G &= c^2 \brac{- dt^2 + dX^i dX^i} + c^{-2}\brac{dx dx + dY^M dY^M }\ , \\
    e^{\varphi} &= 1 \ , \\
    C_4 &= c^4 dt \wedge dX^1 \wedge dX^2 \wedge dX^3 \ ,
\end{align}
\end{subequations}
so as in the previous cases we take our coordinates and fields to have the scalings
\begin{subequations}
\begin{align}
    \hat{t} &= c t \ , \\
    \hat{x} &= c^{-1} x \ ,
\end{align}
\end{subequations}
and
\begin{subequations}
\begin{align}
    \hat{X}^i(\hat{t}, \hat{x}) &= c X^i(t,x) \ , \\
    \hat{Y}^M(\hat{t}, \hat{x}) &= c^{-1} Y^M(t,x) \ . 
\end{align}
\end{subequations}
Unlike the other examples we do not need to rescale the Yang-Mills coupling here. The Wess-Zumino term arising from the pull-back of $C_4$ to the D1-brane worldvolume is inherently non-Abelian \cite{Myers:1999ps} and takes the form
\begin{align} \nonumber
    S_{WZ} &= \frac{i c^4}{2 g_{YM}^2} \tr \int dt dx \, \epsilon_{ijk} D_x X^i [X^j, X^k] \\
    &\equiv \frac{i c^4}{6 g_{YM}^2} \tr \int dt dx \, \partial_x \brac{\epsilon_{ijk} X^i [X^j , X^k]} \ .
\end{align}
Adding this to \eqref{eq: 2d MSYM action} and scaling gives an action with the divergent term
\begin{equation}
    S_{c^4} = - \frac{c^4}{2 g_{YM}^2} \tr \int dt dx \brac{D_x X^i - \frac{i}{2}\epsilon_{ijk} [X^j, X^k]}^2 \ .
\end{equation}
As this is a square we can regulate it with a Hubbard-Stratonovich transformation, yielding
\begin{equation}
    S_{c^4} = \inv{2 g_{YM}^2} \tr \int dt dx \brac{
    G_i \brac{D_x X^i - \frac{i}{2}\epsilon_{ijk} [X^j, X^k]} - \frac{c^{-4}}{4} G_i G_i } \ .
\end{equation}
The action is now finite and the $c\to\infty$ limit can be taken, and we find
\begin{align} \nonumber
    S_{D3NC} = \inv{2g_{YM}^2} \tr \int dt dx \bigg( &
    F_{01}^2 + D_t X^i D_t X^i + G^i \brac{D_x X^i - \frac{i}{2}\epsilon_{ijk} [X^j, X^k]} \\ \label{eq: D1-D3NC action}
    &- D_x Y^M D_x Y^M + [X^i, Y^M]^2 \bigg) \ .
\end{align}
The Hubbard-Stratonovich field becomes a Lagrange multiplier imposing
\begin{equation} \label{eq: Nahm equations}
    D_x X^i = \frac{i}{2}\epsilon_{ijk} [X^j, X^k] \ ,
\end{equation}
which we recognise as Nahm's equations. The story then goes through as before, with the first two terms defining a quantum-mechanical theory on the moduli space of \eqref{eq: Nahm equations} and the terms involving $Y^M$ inducing a potential after we solve the equation of motion
\begin{equation}
    D_x^2 \,Y^M - [X^i,[X^i,Y^M]] = 0 \ .
\end{equation}

In order to understand the relation between this theory and the D1NC limit of $\cal{N}=4$ SYM we will review the string-theoretic embedding of the set-up \cite{Diaconescu:1996rk}. To simplify the discussion we will focus on the case of two D3-branes, {\it i.e.} we work in the $SU(2)$ theory. The physical picture is that the monopoles in the four-dimensional theory are engineered by separating the D3-branes (induced by the field-theoretic VEV of the field $X$) and stretching the D1-branes between them. We should therefore only expect to recover the bound-state description when we place \eqref{eq: D1-D3NC action} on a finite spatial interval.  However, with appropriate boundary conditions (which we discuss in more detail in appendix \ref{sect: appendix moduli space d1-d3}) for the dynamical fields $\{X^i, A_x\}$ at the interval's endpoints \cite{Nahm:1979yw, Hitchin:1983ay} it has been proven that the moduli space of solutions is isometric to the $k$-monopole moduli space of the $SU(2)$ BPS monopole equations \cite{Nakajima:1990} found as the constraint in the D1NC theory. The kinetic terms in \eqref{eq: D1-D3NC action} are therefore equal to those in \eqref{eq: D1NC action} for this case, and one would expect that such a correspondence holds for any choice of gauge-group. In order to show complete equivalence between the theories we also need to show that the potentials induced by non-zero values of $Y^M$ are the same, which we discuss in appendix \ref{sect: appendix moduli space d1-d3}. The extension to larger numbers of D3-branes is technically more complicated but conceptually simple \cite{Hurtubise:1989qy,Diaconescu:1996rk}. We must separate the $N$ D3-branes, leading to $(N-1)$ intervals in which a specified number (that is generically different for each interval) of D1-branes reside. Nahm's equations must then be solved in each interval and appropriate boundary conditions imposed at the D3-brane positions.

To help convince ourselves that the actions are equivalent we can check that the symmetries of the two theories match. The Lagrangian is invariant up to total time-derivative terms under an $SO(2,1)$ transformation group which acts infinitesimally on the coordinates as
\begin{subequations}
\begin{align}
    \hat{t} &= t + f(t) \ , \\
    \hat{x}^i &= x^i(1-\dot{f}) \ , \\
    f(t) &= a + bt + ct^2 \ ,
\end{align}
\end{subequations}
but as these transformations do not leave the interval invariant when $b$ and $c$ are non-zero we see that the only remaining symmetry is temporal translations. Requiring that the finite interval is preserved also kills off the spatial translations and Galilean boost symmetries of the Lagrangian. As the gauge group contains an Abelian factor we see that the R-symmetry for the dynamical scalar fields consists of $SO(3)$ rotations and an Abelian shift symmetry, forming the group $ISO(3)$. In fact, the shift symmetry can also gain linear time-dependence whilst remaining a symmetry of the action. At first glance it appears something similar occurs for the transverse scalars, leading to an $ISO(5)$ factor. However, the shift transformations change the boundary values of the fields and are therefore not symmetries, leaving us with only an $SO(5)$ group. The reason we don't have to worry about this for the dynamical fields is that their boundary conditions fix the residue of poles at the interval's endpoints, and these are unaffected by a finite shift.
Putting this all together, we see that as for the D1NC theory the symmetries of \eqref{eq: D1-D3NC action} are the Galilean group, though now with a novel realisation as a joint set of spacetime and R-symmetry transformations, and the $SO(5)$ rotations of $Y^M$.

\subsection{The Dual Gravitational Picture} \label{sect: d1-d3 gravity}

\subsubsection{The D1NC Limit of the D3-Brane}

We will now examine the two limits from the dual gravitational perspective. The D1NC limit of type IIB supergravity is implemented by making the field redefinition
\begin{subequations} \label{eq: D1NC gravity field redef}
\begin{align}
    \hat{G}_{\mn} &= c^2 \cal{T}_{\mn} + c^{-2} \cal{E}_{\mn} \ , \\
    \hat{G}^{\mn} &= c^2 \cal{E}^{\mn} + c^{-2} \cal{T}^{\mn} \ , \\
    e^{\hat{\varphi}} &= c^{-2} e^{\varphi} \ , \\
    \hat{B}_2 &= B_2 \ , \\
    \hat{C}_0 &= C_0 \ , \\
    \hat{C}_2 &= c^4 e^{-\varphi} \tau^0 \wedge \tau^1 + C_2 \ , \\
    \hat{C}_4 &= C_4 \ ,
\end{align}
\end{subequations}
before taking the $c\to\infty$ limit, where $\cal{T}$ and $\cal{E}$ are defined as in section \ref{sect: General} with veilbeins $\tau^A$ and $e^a$. A full discussion of the action obtained from this limit is given in appendix \ref{sect: appendix d1nc sugra limit}. Using the general discussion in section \ref{sect: General}, the non-zero fields of the smeared D3-D1NC solution are
\begin{subequations} \label{eq: smeared D3-D1NC solution}
\begin{align}
    \hat{G} &= c^2\left[- H_3^{-1/2} dt^2 + H_3^{1/2}  dX^2 \right]+	c^{-2} \left[ H_3^{-1/2} dx^i dx^i + H_3^{1/2} dY^M dY^M \right] \ , \\
    e^{\hat{\varphi}} &= c^{-2} \ , \\
    \hat{C}_2 &= c^4 dt \wedge dX \ , \\
    \hat{F}_5 &= d H_3^{-1} \wedge dt \wedge dx^1 \wedge dx^2 \wedge dx^3 + *\brac{d H_3^{-1} \wedge dt \wedge dx^1 \wedge dx^2 \wedge dx^3 } \ ,
\end{align}
\end{subequations}
with
\begin{equation}
    H_3 = 1 + \frac{R^3}{|Y|^3} \ ,
\end{equation}
which is of the form \eqref{eq: D1NC gravity field redef}. In line with the string theory set-up we take the $X$-coordinate to parameterise a finite interval. Upon taking the near-horizon limit of the D3-brane's harmonic function the metric becomes
\begin{equation}
     \hat{G} = c^2\left[-\left(\frac{|Y|}{R}\right)^{\frac32}dt^2 +\left(\frac{R}{|Y|}\right)^{\frac32} dX^2 \right]+	c^{-2} \left[\left(\frac{|Y|}{R}\right)
    ^{\frac32}dx^i dx^i + \left(\frac{R}{|Y|}\right)^{\frac32}dY^M dY^M \right] \ .
\end{equation}
The radial component of the metric is not scale-invariant, which we can remedy using the Newton-Cartan-Weyl symmetry \eqref{eq: ncw general transformation}. However, as the dilaton scales with $c$ it also transforms; as the dilaton in the gauge \eqref{eq: smeared D3-D1NC solution} is constant any NCW transformation will induce a running dilaton and break conformal invariance. The solution there does not admit a scaling symmetry. The symmetries of the solution are then translations in $t$, the Euclidean group of the $x^i$-plane, and $SO(5)$ rotations of $Y^M$, matching those of the field theory.


\subsubsection{The D3NC Limit of the D1-Brane}

We can similarly work with the D3NC limit using the field redefinitions
\begin{subequations} \label{eq: D3NC gravity field redef}
\begin{align}
    \hat{G}_{\mn} &= c^2 \cal{T}_{\mn} + c^{-2} \cal{E}_{\mn} \ , \\
    \hat{G}^{\mn} &= c^2 \cal{E}^{\mn} + c^{-2} \cal{T}^{\mn} \ , \\
    e^{\hat{\varphi}} &= e^{\varphi} \ , \\
    \hat{B}_2 &= B_2 \ , \\
    \hat{C}_0 &= C_0 \ , \\
    \hat{C}_2 &= C_2 \ , \\
    \hat{C}_4 &= c^4 e^{-\varphi} \tau^0 \wedge \ldots \wedge \tau^3 + C_4 \ .
\end{align}
\end{subequations}
The D1-D3NC solution has the non-zero fields
\begin{subequations}
\begin{align}
    \hat{G} &= c^2 \left[ - H_1^{-1/2} dt^2 + H_1^{1/2} dx^i dx^i \right] + c^{-2} \left[ H_1^{-1/2} dX^2 + H_1^{1/2} dY^M dY^M \right] \ , \\
    e^{\hat{\varphi}} &= H_1^{1/2} \ , \\
    \hat{C}_2 &= \brac{H_1^{-1} - 1 } dt \wedge dX \ , \\
    \hat{C}_4 &= c^4 dt \wedge dx^1 \wedge dx^2 \wedge dx^3 \ ,
\end{align}
\end{subequations}
with the D1-brane harmonic function
\begin{equation}
    H_1 = 1 + \frac{R^3}{|Y|^3} \ ,
\end{equation}
leading to the metric
\begin{equation}
     \hat{G} = c^2\left[-\left(\frac{|Y|}{R}\right)^{\frac32}dt^2 +\left(\frac{R}{|Y|}\right)^{\frac32}dx^i dx^i\right]+	c^{-2} \left[\left(\frac{|Y|}{R}\right)
    ^{\frac32} dX^2  + \left(\frac{R}{|Y|}\right)^{\frac32}dY^M dY^M \right] \ .
\end{equation}
We again take $X$ to be along a finite interval here. Although we can use NCW transformations to render the metric scale-invariant, since the running dilaton is $c$-independent it remains unchanged following the NCW transformtation. Thus, as in the D1NC case, the solution does not admit a scaling symmetry. The symmetries are then exactly as above, as can be seen by noting that the two metrics are identical upon interchanging $X$ and $x^i$.


\section{The M2-M5 System} \label{sect: m2-m5}

The last system we will consider is the bound state of M2-branes and M5-branes with the configuration
\begin{align} \label{eq: M2-M5 brane config}
\begin{array}{rrrrrrrr}
M2: & 0 & 1 & 2 & & & & \\
M5:& 0 & 1 & &3 &4 & 5 & 6 \ \ .
\end{array}
\end{align}
This is conceptually quite similar to the D1-D3 system, with the M2-branes stretched between the M5-branes along the second spatial direction. The novelty here is that we lack a description of the $\cal{N}=(2,0)$ theory on the M5-brane worldvolume we can only compute one of the reciprocal limits. The embedding of the field theory limits within the joint M2NC-M5NC limit of M-theory means we still expect the pair to be equivalent, and so the theory we will find on the M2-branes serves as a prediction for a decoupling limit of the $\cal{N}=(2,0)$ theory.

\subsection{The M5NC Limit of ABJM} \label{sect: ABJM limit}

\subsubsection{The Decoupling Limit}


The simplest way to analyse the M2-M5 system is from the perspective of the M2-branes. If we perform the M2NC limit to isolate the low-energy degrees of freedom and take a $\bb{Z}_k$ orbifold of the system, it is well-known that the system is described by the ABJM theory. Let us first deal with the Bosonic fields and tackle the Fermions later; the theory has two $U(N)$ gauge-fields and four bifundamental complex scalar fields with the action
\begin{align} \nonumber
    \hat{S}_B = \tr \int d^2\hat{\sigma} d\hat{x} \Bigg[& - \hat{D}_{\mu} \hat{\bcalz}_P \hat{D}^{\mu} \hat{\calz}^P + \frac{k \varepsilon^{\mn\rho}}{4\pi} \bigg( \hat{A}_{\mu}^L \hat{\partial}_{\nu} \hat{A}_{\rho}^L - \frac{2i}{3} \hat{A}_{\mu}^L \hat{A}_{\nu}^L \hat{A}_{\rho}^L - \hat{A}_{\mu}^R \hat{\partial}_{\nu} \hat{A}_{\rho}^R \\ \label{eq: ABJM action}
    &+ \frac{2i}{3} \hat{A}_{\mu}^R \hat{A}_{\nu}^R \hat{A}_{\rho}^R \bigg) - \frac{8\pi^2}{3k^2 } \hat{\bar{\Upsilon}}^{R}_{PQ} \hat{\Upsilon}^{PQ}_{R} \Bigg] \ ,
\end{align}
where we are using the notation
\begin{subequations}
\begin{align}
    \hat{\Upsilon}^{PQ}_{R} &= [\hat{\calz}^P , \hat{\calz}^Q ; \hat{\bcalz}_R] - \inv{2} \delta^P_R [ \hat{\calz}^S , \hat{\calz}^Q; \hat{\bcalz}_S] + \inv{2} \delta^Q_R [ \hat{\calz}^S , \hat{\calz}^P ; \hat{\bcalz}_S]\ , \\
    [\hat{\calz}^P , \hat{\calz}^Q ; \hat{\bcalz}_R] &= \hat{\calz}^P \hat{\bcalz}_R \hat{\calz}^Q - \hat{\calz}^Q \hat{\bcalz}_R \hat{\calz}^P \ .
\end{align}
\end{subequations}
The M5-brane supergravity solution is
\begin{subequations}
\begin{align}
    ds^2 &= H_5^{-1/3} \brac{\eta_{\alpha\beta} d\sigma^{\alpha}d\sigma^{\beta} + dX^i dX^i} + H_5^{2/3} \brac{dx^2 + dY^M dY^M} \  \\
    C_6 &= \brac{H_5^{-1} - 1} d\sigma^0 \wedge d\sigma^1 \wedge dX^2 \wedge \ldots \wedge dX^5 \ .
\end{align}
\end{subequations}
The M5NC limit of a flat background is found using the harmonic function
\begin{equation}
    H_5 = c^{-3} \ ,
\end{equation}
giving the solution
\begin{subequations}
\begin{align}
    ds^2 &= c \brac{\eta_{\alpha\beta} d\sigma^{\alpha}d\sigma^{\beta} + dX^i dX^i} + c^{-2} \brac{dx^2 + dY^M dY^M} \  \\ \label{eq: M5NC C6}
    C_6 &= c^3 d\sigma^0 \wedge d\sigma^1 \wedge dX^2 \wedge \ldots \wedge dX^5 \ ,
\end{align}
\end{subequations}
before taking the limit $c\to\infty$, where we have dropped a subleading term in $C_6$. Combining $X^i$ into two complex coordinates $\calz^I$ and $Y^M$ into two coordinates $\calz^A$, we see that we must rescale the ABJM coordinates and fields with the powers
\begin{subequations}
\begin{align}
    \hat{\sigma}^{\alpha} &= c^{1/2} \sigma^{\alpha} \ , \\
    \hat{x} &= c^{-1} x \ ,
\end{align}
\end{subequations}
and
\begin{subequations}
\begin{align}
    \hcalz^I(\hat{\sigma}, \hat{x}) &= c^{1/2} \calz^I(\sigma, x) \ , \\
    \hcalz^A(\hat{\sigma} , \hat{x}) &= c^{-1} \calz^A(\sigma, x) \ , \\
    \hat{A}_{\alpha}^{L/R}(\hat{\sigma}, \hat{x}) &= c^{-1/2} A_{\alpha}^{L/R}(\sigma,x) \ , \\
    \hat{A}_{x}^{L/R}(\hat{\sigma}, \hat{x}) &= c A_{x}^{L/R}(\sigma,x) \ .
\end{align}
\end{subequations}
As usual, the scaling of the components of the two gauge fields is fixed by requiring that gauge-invariance is maintained throughout the limit. 

We can now implement the M5NC of ABJM. Since the Chern-Simons terms are topological they are unchanged by the scaling, so we will focus on the scalar sector of \eqref{eq: ABJM action}. After rescaling this takes the form 
\begin{subequations}
\begin{align}
    \hat{S}_{\calz} &= c^3 S_+ + S_0 + O(c^{-3}) \ , \\
    S_+ &= - \tr \int d^2\sigma dx \brac{D_x \bcalz_I D_x \calz^I - \frac{4\pi^2}{k^2} [\bcalz_I, \bcalz_K ; \calz^I] [\calz^J , \calz^K ; \bcalz_J]} \ , \\ \nonumber
    S_0 &= - \tr \int d^2\sigma dx \bigg(
    D_{\alpha} \bcalz_I D^{\alpha} \calz^I + D_x \bcalz_A D_x \calz^A - \frac{4\pi^2}{3k^2} \Big( 
    4[\bcalz_I, \bcalz_A ; \calz^J] [\calz^I , \calz^A ; \bcalz_J] \\ \nonumber
    &\hspace{3.0cm} + 2 [\bcalz_I , \bcalz_J ; \calz^A] [ \calz^I , \calz^J ; \bcalz_A] - [\bcalz_I , \bcalz_A ; \calz^I] [ \calz^J , \calz^A ; \bcalz_J] \\
    &\hspace{3.0cm} - [\bcalz_A , \bcalz_I ; \calz^A] [ \calz^J , \calz^I ; \bcalz_J] - [\bcalz_J , \bcalz_I ; \calz^J] [ \calz^A , \calz^I ; \bcalz_A]
    \Big) \bigg) \ .
\end{align}
\end{subequations}
As in the previous sections the action has terms that diverge as we take $c \to \infty$. However, using the identity
\begin{equation}
    \partial_x \tr \brac{\bcalz_I [\calz^I , \calz^J ; \bcalz_J]} = 2\tr \brac{D_x \bcalz_I [\calz^I , \calz^J ; \bcalz_J] - [\bcalz_I , \bcalz_J ; \calz^J] D_x \calz^I} \ ,
\end{equation}
we see that $S_+$ can be rewritten in the form
\begin{align}
    S_+ = - \tr \int d^2\sigma dx \brac{
    \abs{ D_x \calz^I - \frac{2\pi}{k} [\calz^I , \calz^J ; \bcalz_J] }^2 + \frac{\pi}{k} \partial_x \brac{\bcalz_I [\calz^I , \calz^J ; \bcalz_J] } } \ .
\end{align}
The second term is a total derivative and therefore does not contribute to the dynamics of the theory; however, it will contribute a divergent term to the energy of any non-trivial state in the theory. In the previous example we found that such terms are cancelled by the contribution of the constant divergent spacetime $(p+1)$-form field arising from the $p$-brane Newton-Cartan limit. In this case, we would therefore expect the term to be cancelled by the coupling of the M2-branes to the divergent flat six-form field \eqref{eq: M5NC C6}. It is clear that using the M-theory analogue of the non-Abelian D-brane Wess-Zumino terms gives a contribution proportional to the total derivative term, and so with an appropriate normalisation the two cancel; we propose that this is the case here. The remaining divergent term is a square so can be regulated through the introduction of a Hubbard-Stratonovich field $H^I$. This allows us to take the $c\to\infty$ limit, leaving us with the scalar action
\begin{align} \nonumber
    S_{\calz} = - \tr \int d^2\sigma dx \bigg(&
    D_{\alpha} \bcalz_I D^{\alpha} \calz^I + D_x \bcalz_A D_x \calz^A + \Bar{H}_I \brac{D_x \calz^I - \frac{2\pi}{k} [\calz^I , \calz^J ; \bcalz_J]}   \\ \nonumber
    &+\brac{D_x \bcalz_I + \frac{2\pi}{k} [\bcalz_I , \bcalz_J ; \calz^J]} H^I - \frac{4\pi^2}{3k^2} \Big( 
    4[\bcalz_I, \bcalz_A ; \calz^J] [\calz^I , \calz^A ; \bcalz_J] 
     \\ \nonumber
    &+ 2 [\bcalz_I , \bcalz_J ; \calz^A] [ \calz^I , \calz^J ; \bcalz_A]  - [\bcalz_I , \bcalz_A ; \calz^I] [ \calz^J , \calz^A ; \bcalz_J] \\ \label{eq: M2-M5NC Scalar Action}
    &- [\bcalz_A , \bcalz_I ; \calz^A] [ \calz^J , \calz^I ; \bcalz_J] - [\bcalz_J , \bcalz_I ; \calz^J] [ \calz^A , \calz^I ; \bcalz_A]
    \Big) \bigg) \ .
\end{align}

The constraint imposed by $H^i$ after taking the limit is
\begin{equation} \label{eq: ABJM BPS equations}
    D_x \calz^I - \frac{2\pi}{k} [\calz^I , \calz^J ; \bcalz_J] = 0 \ .
\end{equation}
This equation was found in \cite{Terashima:2008sy} using the constraints imposed for half-BPS field configurations, and was interpreted as the ABJM realisation of the Basu-Harvey equations describing M2-branes ending on M5-branes \cite{Basu:2004ed}. The remaining scalar fields $\calz^A$ have become non-dynamical after the limit, with their role being to introduce a potential for the $\calz^I$ fields after imposing their equations of motion
\begin{align} \nonumber
    0 = \ & D_x^2 \calz^A + \frac{4\pi^2}{3k^2} \bigg( 
    4 [[\calz^I, \calz^A ; \bcalz_J], \calz^J ; \bcalz_I] - 2 [\calz^I , \calz^J ; [\bcalz_I , \bcalz_J ; \calz^A]] \\
    &- [[\calz^J , \calz^A ; \bcalz_J], \calz^I ; \bcalz_I] - [[\calz^I , \calz^J ; \bcalz_I] , \calz^A ; \bcalz_I] + [\calz^I , \calz^A ; [\bcalz_I , \bcalz_J ; \calz^J]] \bigg) \ ,
\end{align}
leading to the contribution
\begin{equation}
    S_V = - \tr \int d^2\sigma dx \, \partial_x \brac{
    \bcalz_A D_x \calz^A } \ ,
\end{equation}
to the action.

So far we have ignored the Fermions in our analysis. We have not yet found a compelling way of predicting how the Fermionic fields should scale in order for the action to remain supersymmetric, so we will proceed using some guiding principles and some trial-and-error. The relativistic theory has four complex 2-component spinor fields $\hat{\Psi}_P$\footnote{The relativistic Fermionic action is given in \cite{Bagger:2012jb}.}, whose index we can again split into $I$ and $A$. The BPS equations \eqref{eq: ABJM BPS equations} are found in the relativistic theory by requiring that the supersymmetry transformation parameter is an eigenspinor of $\gamma^2$, so we must split our 2-component spinor fields $\hat{\Psi}_I(x)$ and $\hat{\Psi}_A(x)$ similarly, taking
\begin{equation}
    \hat{\Psi}_{I/A}(\hat{\sigma},\hat{x}) = \hpsi_{I/A}^+(\hat{\sigma},\hat{x}) \epsilon_{+} + \hpsi_{I/A}^-(\hat{\sigma},\hat{x}) \epsilon_- \ ,
\end{equation}
where $\hpsi_{I/A}^{\pm}(\hat{\sigma},\hat{x})$ are complex Grassmann-valued fields and $\epsilon_{\pm}$ are the normalised eigenspinors of $\gamma^2$ with eigenvalues $\pm1$. In order to take the M5NC limit of the action we will require that the rescaled action has no terms that diverge as we send $c\to\infty$, that each field features in at least one derivative term, and that the R-symmetry that remains unbroken in the Bosonic limit is also preserved here. These are satisfied by the rescaling
\begin{subequations}
\begin{align}
    \hpsi^+_I(\hat{\sigma}, \hat{x}) &= c^{1/4} \psi_I^+(\sigma,x) \ , \\
    \hpsi^-_I(\hat{\sigma}, \hat{x}) &= c^{-5/4} \psi_I^-(\sigma,x) \ , \\
    \hpsi^+_A(\hat{\sigma}, \hat{x}) &= c^{-5/4} \psi_A^+(\sigma,x) \ , \\
    \hpsi^-_A(\hat{\sigma}, \hat{x}) &= c^{1/4} \psi_A^-(\sigma,x) \ ,
\end{align}
\end{subequations}
in terms of which the Fermionic action becomes
\begin{align} \nonumber
    S_{F} = i \tr \int d^2\sigma dx \bigg(& 2\bpsi^{I,+} D_+ \psi_I^+ + 2 \bpsi^{A,-} D_- \psi_A^- + \bpsi^{I,+} D_x \psi_I^- + \bpsi^{I,-} D_x \psi_I^+ \\ \nonumber
    &+ \bpsi^{A,+} D_x \psi_A^- + \bpsi^{A,-} D_x \psi_A^+  +\frac{2\pi}{k} \bigg[ \bpsi^{I,-} [\psi_I^+, \calz^J ; \bcalz_J] \\ \nonumber
    & - \bpsi^{I,+} [\psi_I^-, \calz^J ; \bcalz_J] + \bpsi^{A,-} [\psi_A^+, \calz^J ; \bcalz_J] - \bpsi^{A,+} [\psi_A^-, \calz^J ; \bcalz_J] \\ \nonumber
    & + 2 \bpsi^{I,+} [\psi_J^- , \calz^J ; \bcalz_I] - 2 \bpsi^{I,-} [\psi_J^+ , \calz^J ; \bcalz_I] + 2 \bpsi^{I,+} [\psi_A^- , \calz^A ; \bcalz_I] \\ \nonumber
    &- 2 \bpsi^{A,-} [\psi_I^+, \calz^I ; \bcalz_A] + \epsilon_{AB} \epsilon_{IJ} \Big( 
    \bpsi^{A,+} [\calz^I , \calz^J ; \bpsi^{B,-}] \\ \nonumber
    &- 2 \bpsi^{I,+} [\calz^J , \calz^A ; \bpsi^{B,-}] \Big)  - \epsilon^{AB} \epsilon^{IJ} \Big( \psi_A^+ [\bcalz_I, \bcalz_J ; \psi_B^-] \\
    &- 2 \psi_I^+ [\bcalz_J , \bcalz_A ; \psi_B^-] \Big)  \bigg] \bigg) \ ,
\end{align}
in the $c \to \infty$ limit. We have used
\begin{equation}
    \sigma^{\pm} = \sigma^0 \pm \sigma^1
\end{equation}
as the normalisation of our lightcone coordinates. While this isn't the only scaling we could choose that is consistent with our requirements, it will turn out to be the one that preserves supersymmetry. Note that the limit has conspired to remove the terms quadratic in $\psi_I^-$ and $\psi_A^+$ from the action; they become the Fermionic superpartners of the Lagrange multiplier field $H^I$, imposing the constraints
\begin{subequations}
\begin{align}
    D_x \psi_I^+ + \frac{2\pi}{k} \brac{[\psi_I^+, \calz^J ; \bcalz_J] - 2 [\psi_J^+, \calz^J ; \bcalz_I]} &= 0 \ , \\
    D_x \psi_A^- + \frac{2\pi}{k} \brac{\epsilon_{AB} \epsilon_{IJ} [\calz^I , \calz^J ; \bpsi^{B,-}] - [\psi_A^- , \calz^J ; \bcalz_J]} &= 0 \ .
\end{align}
\end{subequations}
The total action of the M5NC theory is then
\begin{equation} \label{eq: M2-M5NC action}
    S_{M5NC} = S_{CS,k}^L + S_{CS,-k}^R + S_{\calz} + S_F \ .
\end{equation}
In order to see the M-theory interpretation of the limit, let us focus on the bound state of $N$ M2-branes with two M5-branes. As the system we are interested in describing is then obtained by separating the M5-branes along the $x$-direction and stretching the M2-branes between them we should consider \eqref{eq: M2-M5NC action} on a finite interval in order to recover the correct physics. This is completely analogous to the D1-D3 system and so it seems natural that the extension to higher numbers of M5-branes proceeds in the same way. In other words, we must separate the $k$ M5-branes and stretch M2-branes across the $(k-1)$ intervals, with appropriate boundary conditions imposed on the M2-brane fields at the M5-brane positions, i.e. where two intervals meet.

As we are interested in placing the theory on the interval $x\in [-L,L]$, and as such we need to specify boundary conditions for the Bosonic fields at the endpoints. To make contact with the M-theory embedding we will require that these preserve the 'physical' half of the supersymmetry, which will be discussed in section \ref{sect: m2-m5 symmetries}. Supersymmetry-preserving boundary conditions in ABJM were considered in \cite{Chu:2009ms, Berman:2009xd}, where it was found that the constraint \eqref{eq: ABJM BPS equations} preserves half of ABJM's supersymmetry when imposed as a boundary condition on the fields $\calz^I$ and $A_x^{L/R}$. There is, however, some debate over the correct choice of boundary conditions for the components of the gauge fields in the boundary directions. The approach considered in \cite{Chu:2009ms} involves adding extra degrees of freedom to the boundary in the form of a WZW model that couples to the gauge fields in order to cancel the gauge-anomaly, while \cite{Berman:2009xd} proposes that the condition $A_{\pm}^L = A_{\pm}^R$ should be imposed on the boundary. In the D-brane examples considered above the bound state dynamics arose only from a $\sigma$-model on the BPS constraint's moduli space. Here we will for simplicity assume that the same happens here and will not include additional dynamical degrees of freedom at the boundary, {\it i.e.} adopting the second proposal. This may need to be revisited but we do not think it will affect our discussion here.

\subsubsection{Symmetries of the M2-M5 system} \label{sect: m2-m5 symmetries}

As the action \eqref{eq: M2-M5NC action} has not been constructed before we would like to understand its symmetries. We will start by considering the spacetime and R-symmetries before moving on to discuss the supersymmetries.  After taking the limit, the spacetime symmetry of the original relativistic theory splits into parts associated with $\sigma^{\pm}$ and $x$. We see that the symmetries of the 'large' 2d Lorentzian spacetime undergo an enhancement to the conformal transformations
\begin{subequations}
\begin{align}
    \hat{\sigma}^{\pm} &= f^{\pm}(\sigma^{\pm}) \ , \\
    \hat{x} &= x \ ,
\end{align}
\end{subequations}
with the action invariant under the field transformations
\begin{subequations}
\begin{align}
    \hcalz^I(\hat{\sigma}, \hat{x}) &= \calz^I(\sigma, x) \ , \\
    \hcalz^A (\hat{\sigma}, \hat{x}) &= \brac{\partial_+ f^+ \partial_- f^-}^{-1/2} \calz^A(\sigma, x) \ , \\
    \hat{H}^I(\hat{\sigma}, \hat{x}) &= \brac{\partial_+ f^+ \partial_- f^-}^{-1} H^I(\sigma, x) \ , \\
    \hat{A}_{\pm}^{L/R} (\hat{\sigma}, \hat{x}) &= \brac{\partial_{\pm} f^{\pm}}^{-1} A_{\pm}^{L/R}(\sigma,x) \ , \\
    \hat{A}_x^{L/R}(\hat{\sigma}, \hat{x}) &= A_x^{L/R}(\sigma, x) \ , \\
    \hpsi_I^+(\hat{\sigma}, \hat{x}) &= \brac{\partial_- f^-}^{-1/2} \psi_I^+(\sigma, x ) \ , \\
    \hpsi_I^-(\hat{\sigma}, \hat{x}) &= \brac{\partial_- f^-}^{-1/2} \brac{\partial_+ f^+}^{-1} \psi_I^-(\sigma, x ) \ , \\
    \hpsi_A^-(\hat{\sigma}, \hat{x}) &= \brac{\partial_+ f^+}^{-1/2} \psi_A^-(\sigma, x ) \ , \\
    \hpsi_A^+(\hat{\sigma}, \hat{x}) &= \brac{\partial_+ f^+}^{-1/2} \brac{\partial_- f^-}^{-1} \psi_A^+(\sigma, x ) \ .
\end{align}
\end{subequations}

As the Lagrange multiplier imposes a constraint that fixes the $x$-profile of our dynamical fields, the discussion in section \ref{sect: d1-d3} leads us to the conclusion that the translational symmetry in this coordinate must exhibit a $\sigma$-dependent quasi-symmetry enhancement. Working infinitesimally, the coordinate transformation
\begin{subequations}
\begin{align}
    \hat{\sigma}^{\pm} &= \sigma^{\pm} \ , \\
    \hat{x} &= x + a(\sigma) \ , 
\end{align}
\end{subequations}
combined with the field transformations
\begin{subequations}
\begin{align}
    \hcalz^I(\hat{\sigma}, \hat{x}) &= \calz^I(\sigma,x) \ , \\
    \hcalz^A(\hat{\sigma}, \hat{x}) &= \calz^A(\sigma, x) \ , \\
    \hat{H}^I(\hat{\sigma}, \hat{x}) &= \brac{H^I + \partial_{\alpha} a \, D^{\alpha}\calz^I + \inv{4} \partial^2 a\,  \calz^I}(\sigma,x) \ , \\
    \hat{A}_{\pm}^{L/R}(\hat{\sigma}, \hat{x}) &= \brac{A_{\pm}^{L/R} - \partial_{\pm}a \, A_x^{L/R}}(\sigma,x) \ , \\
    \hat{A}_x^{L/R}(\hat{\sigma}, \hat{x}) &= A_x^{L/R}(\sigma,x) \ , \\
    \hpsi_I^+(\hat{\sigma}, \hat{x}) &= \psi_I^+(\sigma,x) \ , \\
    \hpsi_I^-(\hat{\sigma}, \hat{x}) &= \brac{\psi_I^- + \partial_+ a \, \psi_I^+}(\sigma,x) \ , \\
    \hpsi_A^- (\hat{\sigma}, \hat{x}) &= \psi_A^-(\sigma,x) \ , \\
    \hpsi_A^+(\hat{\sigma}, \hat{x}) &= \brac{\psi_A^+ + \partial_- a \, \psi_A^-}(\sigma,x) \ ,
\end{align}
\end{subequations}
are a quasi-symmetry of the action; dropping total derivatives in the $\sigma$-directions\footnote{In other words, we assume the only boundary of our theory occurs in the $x$-coordinate.}, we have invariance of the Lagrangian up to the boundary term
\begin{equation}
    \delta \cal{L} = - \inv{4} \partial_x \tr\brac{\partial^2 a \bcalz_I \calz^I} \ .
\end{equation}
We note that if we take
\begin{equation} \label{eq: a lightcone decomp}
    a = a_+(\sigma^+) + a_-(\sigma^-)
\end{equation}
this term vanishes. If we integrate out $H^I$ and works directly on the constraint surface we find that \eqref{eq: a lightcone decomp} remains a symmetry of the Lagrangian.

This is not the only symmetry involving $x$; we also have a scaling symmetry, which cannot be given arbitrary $\sigma$-dependence but can be taken to have a left-moving and right-moving part. Its infinitesimal form is
\begin{subequations}
\begin{align}
    \hat{\sigma}^{\pm} &= \sigma^{\pm} \ , \\
    \hat{x} &= \brac{ 1 + b_{\pm}(\sigma^{\pm}) } x \ ,
\end{align}
\end{subequations}
with the fields transformations
\begin{subequations}
\begin{align}
    \hcalz^I(\hat{\sigma}, \hat{x}) &= \brac{1- \inv{2}b_{\pm}} \calz^I (\sigma,x) \ , \\
    \hcalz^A(\hat{\sigma}, \hat{x}) &= \brac{1+ \inv{2}b_{\pm}} \calz^A (\sigma,x) \ , \\
    \hat{H}^I(\hat{\sigma}, \hat{x}) &= \brac{\brac{1 + \inv{2}b_{\pm}} H^I - 2 x b_{\pm}' D_{\mp} \calz^I}(\sigma,x) \ , \\
    \hat{A}_{\pm}^{L/R}(\hat{\sigma}, \hat{x}) &= \brac{A_{\pm}^{L/R} - x b_{\pm}' A_x^{L/R}}(\sigma,x) \ , \\
    \hat{A}_{\mp}^{L/R}(\hat{\sigma}, \hat{x}) &= A_{\mp}^{L/R}(\sigma,x) \ , \\
    \hat{A}_x^{L/R}(\hat{\sigma}, \hat{x}) &= \brac{1 - b_{\pm}} A_x^{L/R}(\sigma,x) \ , \\
    \hpsi_I^+(\hat{\sigma}, \hat{x}) &= \brac{1 - \inv{2}b_{\pm}} \psi_I^+(\sigma,x) \ , \\
    \hpsi_A^-(\hat{\sigma}, \hat{x}) &= \brac{1 - \inv{2}b_{\pm}} \psi_A^-(\sigma,x) \ ,
\end{align}
\end{subequations}
for both $b_{\pm}(\sigma^{\pm})$, and the further transformations
\begin{subequations}
\begin{align}
    \hpsi_I^-(\hat{\sigma}, \hat{x}) &= \brac{\brac{1 + \inv{2}b_+} \psi_I^- + x b_+' \psi_I^+}(\sigma, x) \ , \\
    \hpsi_A^+(\hat{\sigma}, \hat{x}) &= \brac{1 + \inv{2}b_+} \psi_A^+(\sigma, x) \ ,
\end{align}
\end{subequations}
for $b_+$, and
\begin{subequations}
\begin{align}
    \hpsi_I^-(\hat{\sigma}, \hat{x}) &= \brac{1 + \inv{2}b_-} \psi_I^-(\sigma, x) \ , \\
    \hpsi_A^+(\hat{\sigma}, \hat{x}) &= \brac{ \brac{1 + \inv{2}b_-} \psi_A^+ + x b_-' \psi_A^-}(\sigma, x) \ ,
\end{align}
\end{subequations}
for $b_-$. We note that if the theory is placed on a finite interval then the both the translation and scaling change the endpoints and are therefore not symmetries of the action. However, if we were to instead work on the half-line $(-\infty,0]$ the scaling symmetry is retained, with the translational symmetry still absent. In both cases, the fact that this transformation is a symmetry up to boundary terms allows us to disentangle the transformations of $\sigma^{\alpha}$ and $x$, which ensures that the two-dimensional conformal symmetry survives even in the presence of the $x$-boundary.

We next deal with the R-symmetry of the theory. This is split by the limit into two groups, each with the same $SU(2)\times U(1)$ structure. The field transformations for the symmetries are
\begin{subequations}
\begin{align}
    \hcalz^I &= \tensor{\cal{U}}{^I_J} \calz^J \ , \\
    \hat{H}^I &= \tensor{\cal{U}}{^I_J} H^J \ , \\
    \hpsi_I^{\pm} &= \tensor{\bar{\cal{U}}}{_I^J} \psi_J^{\pm} \ ,
\end{align}
\end{subequations}
with $\cal{U}\in SU(2)$, and
\begin{subequations}
\begin{align}
    \hcalz^I &= e^{i\theta} \calz^I \ , \\
    \hat{H}^I &= e^{i\theta} H^I \ , \\
    \hpsi_I^{\pm} &= e^{-i\theta} \psi_I^{\pm} \ ,
\end{align}
\end{subequations}
for the first group, and
\begin{subequations}
\begin{align}
    \hcalz^A &= \tensor{\cal{R}}{^A_B} \calz^B \ , \\
    \hpsi_A^{\pm} &= \tensor{\bar{\cal{R}}}{_A^B} \psi_B^{\pm} \ ,
\end{align}
\end{subequations}
with $\cal{R}\in SU(2)$, and
\begin{subequations}
\begin{align}
    \hcalz^A &= e^{i\varphi} \calz^A \ , \\
    \hpsi_A^{\pm} &= e^{-i\varphi} \psi_A^{\pm} \ ,
\end{align}
\end{subequations}
for the second. This pattern of symmetry breaking can be understood from the M-theory setup. Taking the M5NC limit splits the directions transverse to the M2-brane into two equal groups; this splits the $S^7$ associated with rotations of these coordinates into two $S^3$ factors, corresponding to a symmetry breaking pattern $SO(8) \to SO(4) \times SO(4)$. Using the Hopf map to write these as an $S^1$ fibre over $S^2$, the action of the ABJM orbifold is to reduce the periodicity of the fibre. This further breaks the symmetry as $SO(4) \to SU(2) \times U(1)$ for each factor, resulting in the observed R-symmetry.

Next we turn to the supersymmetry of the theory. In line with our treatment of the Fermions in section \ref{sect: ABJM limit}, we split the supersymmetry parameters of the relativistic theory into chiral components with respect to $\gamma^2$. When combined with the R-symmetry indices induced by the non-relativistic limit we are left with complex Grassmann-valued parameters $\halpha_{\pm}$ and $\hxi_{IA,\pm}$, where $\xi_{IA,\pm}$ satisfy the reality condition
\begin{equation}
    (\hxi_{IA,\pm})^* = - \epsilon^{IJ} \epsilon^{AB} \hxi_{JB,\pm} \ .
\end{equation}
If we take the relativistic parameters to scale with $c$ as
\begin{subequations}
\begin{align}
    \halpha_+ &= c^{-5/4} \alpha_+ \ , \\
    \halpha_- &= c^{1/4} \alpha_- \ , \\
    \hxi_{IA,+} &= c^{1/4} \xi_{IA,+} \ , \\
    \hxi_{IA,-} &= c^{-5/4} \xi_{IA,-} \ ,
\end{align}
\end{subequations}
then we find that the supersymmetry transformations for the Bosonic fields are finite in $c$, and the only divergences that appear in the transformations of the Fermions are proportional to the constraint \eqref{eq: ABJM BPS equations}. This (combined with the fact that the only divergent term in the action was a squared quantity) means we can apply the algorithm devised in \cite{Lambert:2019nti} for constructing the supersymmetry transformations after the limit has been taken. There is therefore a symmetry in the non-relativistic theory for each of the four parameters $\{\alpha_{\pm}, \xi_{IA,\pm}\}$. 

However, this is not the full story. We saw above that translational symmetry in the $x$ coordinate could be made arbitrarily dependent on the $\sigma$ coordinates while only changing the Lagrangian by boundary terms; since the transformations parameterised by $\alpha_+$ and $\xi_{IA,-}$ square onto this transformation we expect that these transformations should also be extended to $\sigma$-dependent symmetries, as occurs in \cite{Lambert:2024yjk}. It is not too hard to show that with a minor modification of the transformations this is the case: explicitly, the action is invariant under the transformations
\begin{subequations}
\begin{align}
    \delta \calz^I &= 0 \ , \\
    \delta \calz^A &= i \epsilon^{AB} \alpha_+ \psi_B^- \ , \\
    \delta A_+^L &= \frac{2\pi}{k} \bigg( 
    \epsilon^{IJ} \balpha_+ \psi_-^+ \bcalz_J - \epsilon_{IJ} \alpha_+ \calz^J \bpsi^{I,+} \bigg) \ , \\
    \delta A_-^L &= 0 \ , \\
    \delta A_x^L &= 0 \ , \\ \nonumber
    \delta H^I &= i \epsilon^{IJ} \balpha_+ D_x \psi_J^- - \frac{2\pi i}{k} \bigg( \balpha_+ \epsilon_{AB} [\calz^B , \calz^I ; \bpsi^{A,-}] + \balpha_+ \epsilon^{IJ} [\psi_J^- , \calz^K ; \bcalz_K] \\
    & \quad - \alpha_+ \epsilon^{AB} [\psi_A^- , \calz^I ; \bcalz_B] \bigg) - i \partial_+ \balpha_+ \epsilon^{IJ} \psi_J^+ \ , \\
    \delta \psi_I^+ &= - \brac{D_x \calz^J + \frac{2\pi}{k} [\calz^J , \calz^K ; \bcalz_K]} \epsilon_{IJ} \alpha_+ \ , \\
    \delta \psi_I^- &= 2 D_- \calz^J \epsilon_{IJ} \alpha_+ \ , \\
    \delta \psi_A^+ &= \frac{2\pi}{k} [\calz^I , \calz^J ; \bcalz_A] \epsilon_{IJ} \alpha_+ - \brac{D_x \calz^B - \frac{2\pi}{k} [\calz^B , \calz^I ; \bcalz_I] } \epsilon_{AB} \balpha_+ \ , \\
    \delta \psi_A^- &= 0 \ ,
\end{align}
\end{subequations}
and
\begin{subequations}
\begin{align}
    \delta \calz^I &= 0 \ , \\
    \delta \calz^A &= i \bxi^{IA}_- \psi_I^+ \ , \\
    \delta A_+^L &= 0 \ , \\
    \delta A_-^L &= - \frac{2\pi}{k} \brac{ \bxi^{IA}_- \psi_A^- \bcalz_I - \xi_{IA,-} \calz^I \bpsi^{A,-} } \ , \\
    \delta A_x^L &= 0 \ , \\ \nonumber
    \delta H^I &= - i \bxi^{IA}_- D_x \psi_A^+ + \frac{2\pi i}{k} \bxi_-^{IA} [\psi_A^+ , \calz^J ; \bcalz_J] + \frac{2\pi i}{k} \bxi_-^{JA} \Big( [\psi_J^+ , \calz^I ; \bcalz_A] \\
    &\quad - 2 [\psi_A^+ , \calz^I ; \bcalz_J] \Big) - \frac{2\pi i}{k} \xi_{JA,-} [\calz^A , \calz^J ; \bpsi^{J,+}] + i \partial_- \bxi^{IA}_- \psi_A^- \ , \\
    \delta \psi_I^+ &= 0 \ , \\
    \delta \psi_I^- &=  \brac{D_x \calz^A + \frac{2\pi}{k} [\calz^A , \calz^K ; \bcalz_K]} \xi_{IA,-} + \frac{4\pi}{k} [\calz^J , \calz^A ; \bcalz_I] \xi_{JA,-} \ , \\
    \delta \psi_A^+ &= 2 D_+ \calz^I \xi_{IA,-} \ , \\
    \delta \psi_A^- &= - \brac{D_x \calz^I + \frac{2\pi}{k} [\calz^I , \calz^J ; \bcalz_J]} \xi_{IA,-} \ ,
\end{align}
\end{subequations}
for any $\sigma$-dependent parameters $\alpha_+(\sigma)$ and $\xi_{IA,-}(\sigma)$. The transformations for $A_{\mu}^R$ are identical to those of $A_{\mu}^L$ with the ordering of the fields swapped. Similarly, the symmetries of the $\sigma$ coordinates are enhanced to the 2d conformal algebra. As the transformations with parameters $\alpha_-$ and $\xi_{IA,+}$ square onto these transformations it is reasonable to expect that these symmetries are enhanced to the 2d superconformal algebra. We again find that after a small modification of the transformations this is true. Taking $\alpha_- = \alpha_-(\sigma^-)$, we have a symmetry of the action under the transformations
\begin{subequations}
\begin{align}
    \delta \calz^I &= -i \epsilon^{IJ} \balpha_- \psi_J^+ \ , \\
    \delta \calz^A &= - i \epsilon^{AB} \alpha_- \psi_B^+ \ , \\
    \delta A_+^L &= 0 \ , \\
    \delta A_-^L &= \frac{2\pi}{k} \brac{
    \balpha_- \brac{\epsilon^{IJ} \psi_I^- \bcalz_J - \epsilon_{AB} \calz^B \bpsi^{A,-}} - \alpha_- \brac{\epsilon_{IJ} \calz^J \bpsi^{I,-} - \epsilon^{AB} \psi_A^- \bcalz_B} } \ , \\
    \delta A_x^L &= - \frac{2\pi}{k} \brac{\epsilon^{IJ} \balpha_- \psi_I^+ \bcalz_J - \epsilon_{IJ} \alpha_- \calz^J \bpsi^{I,+}} \ , \\ \nonumber
    \delta H^I &= \epsilon^{IJ} \balpha_- \brac{2i D_- \psi_J^- + \frac{2\pi i}{k} [\psi_J^+ , \calz^A ; \bcalz_A] - \frac{4\pi i}{k} [\psi_A^+ , \calz^A ; \bcalz_J]} \\ \nonumber
    &\quad + \frac{2\pi i}{k} \epsilon_{AB} \balpha_- \brac{[\calz^A , \calz^B ; \bpsi^{I,+}] + [\calz^B, \calz^I ; \bpsi^{A,+}] } \\
    &\quad + \frac{2\pi i}{k} \epsilon^{AB} \alpha_- [\psi_A^+ , \calz^I ; \bcalz_B] + 2 i \epsilon^{IJ} \partial_- \balpha_- \psi_J^- \ , \\
    \delta \psi_I^+ &= - 2 D_+ \calz^J \epsilon_{IJ} \alpha_- \ , \\
    \delta \psi_I^- &= \epsilon_{IJ} H^J \alpha_- + \frac{2\pi}{k} [\calz^J , \calz^A ; \bcalz_A] \epsilon_{IJ} \alpha_- + \frac{2\pi}{k} [\calz^A , \calz^B ; \bcalz_I] \epsilon_{AB} \balpha_-  \ , \\
    \delta \psi^+_A &=  - 2 D_+ \calz^B \epsilon_{AB} \balpha_-
    -2 \epsilon_{AB} \calz^B \partial_- \balpha_- \ , \\
    \delta \psi_A^- &= D_x \calz^B \epsilon_{AB} \balpha_- + \frac{2\pi}{k} [\calz^I , \calz^J ; \bcalz_A] \epsilon_{IJ} \alpha_- + \frac{2\pi}{k} [\calz^B , \calz^I ; \bcalz_I] \epsilon_{AB} \balpha_- \ ,
\end{align}
\end{subequations}
and taking $\xi_{IA,+} = \xi_{IA,+}(\sigma^+)$ we have a symmetry for
\begin{subequations}
\begin{align}
    \delta \calz^I &= i \bxi^{IA}_+ \psi_A^- \ , \\
    \delta \calz^A &= -i \bxi^{IA}_+ \psi_I^- \ , \\
    \delta A_+^L &= \frac{2\pi}{k} \brac{
    \bxi_+^{IA} \brac{\psi_I^+ \bcalz_A - \psi_A^+ \bcalz_I} - \xi_{IA,+} \brac{\calz^A \bpsi^{I,+} - \calz^I \bpsi^{A,+}} } \ , \\
    \delta A_-^L &= 0 \ , \\
    \delta A_x^L &= \frac{2\pi}{k} \brac{\bxi_+^{IA} \psi_A^- \bcalz_I - \calz^I \bpsi^{A,-} } \ , \\ \nonumber
    \delta H^I &= - \bxi^{IA}_+ \bigg( 
    2i D_+ \psi_A^+ - \frac{2\pi i}{k} [\psi_A^- , \calz^B ; \bcalz_B] + \frac{4\pi i}{k} [\psi_J^- , \calz^J ; \bcalz_A] + \frac{4\pi i}{k} [\psi_B^- , \calz^B ; \bcalz_A] \\ \nonumber
    & \quad - \frac{4\pi i}{k} \epsilon_{AB} \epsilon_{IJ} [\calz^B , \calz^J ; \bpsi^{I,-}] \bigg) + \frac{2\pi i}{k} \bxi^{JA}_+ \bigg( [\psi_J^- , \calz^I ; \bcalz_A] \\
    & \quad - \epsilon_{JK} \epsilon_{AB} [\calz^B , \calz^I ; \bpsi^{K,+}] \bigg) - 2i \partial_+ \bxi^{IA}_+ \psi_A^+  \ , \\
    \delta \psi_I^+ &= - \brac{D_x \calz^A - \frac{2\pi}{k} [\calz^A , \calz^K ; \bcalz_K]} \xi_{IA,+} + \frac{4\pi}{k} [\calz^J , \calz^A ; \bcalz_I] \xi_{JA,+} \ , \\ \nonumber
    \delta \psi_I^- &= \brac{D_x \calz^A + \frac{2\pi}{k} [\calz^A , \calz^K ; \bcalz_K]} \xi_{IA,-} + \frac{4\pi}{k} [\calz^J , \calz^A ; \bcalz_I] \xi_{JA,-} \\
    &\quad + 2 \calz^A \partial_+ \xi_{IA,+} \ , \\
    \delta \psi_A^+ &= H^I \xi_{IA,+} + \frac{4\pi}{k} [\calz^I , \calz^B ; \bcalz_A] \xi_{IB,+} - \frac{2\pi}{k} [\calz^I , \calz^B ; \bcalz_B] \xi_{IA,+} \ , \\
    \delta \psi_A^- &= - 2 D_- \calz^I \xi_{IA,+} \ .
\end{align}
\end{subequations}

To summarise, we have found that the action \eqref{eq: M2-M5NC action} has the symmetries of a two-dimensional $\cal{N}=(4,2)$ super-conformal field theory. Given that ABJM is expected to have enhanced supersymmetry when $k=1,2$ it seems likely that the same should be true here, giving an $\cal{N}=(4,4)$ SCFT. This is in line with the M-theory picture in which the theory describes the dynamics of self-dual strings on the M5 worldvolume.

\subsection{The M2NC Limit of the Six-Dimensional \texorpdfstring{$\cal{N}=(2,0)$}{N=(2,0)} Theory} \label{sect: MNC of (2,0)}

In each of the previous examples we saw that decoupling limits of string theory associated to bound states of D-branes led us to an equality between the D$q$NC limit of a $(p+1)$-dimensional theory and the reciprocal D$p$NC limit of a $(q+1)$-dimensional theory. We expect the same to hold for the M2-M5 system. Unlike the previous examples, however, we lack a description of the $\cal{N}=(2,0)$ theory, and so we cannot test this directly. In order to get an understanding of the limit we will study the dimensional reduction of the system and see what features can be extracted.

There are two different circle reductions of the brane configuration  \eqref{eq: M2-M5 brane config} that lead to D-branes\footnote{Reducing transverse to both the M2-brane and M5-brane will lead to an intesecting F1 and NS5-brane configuration related to little string theory. But without a description of little string theory that leaves us none the wiser.}. If we reduce on the common spatial direction we obtain the bound state of fundamental strings with D4-branes in the configuration
\begin{align} \label{eq: F1-D4 brane config}
\begin{array}{rrrrrrr}
F1: & 0 &  1 & & & & \\
D4: & 0 &  &2 &3 & 4 & 5 \ \ , \\
\end{array}
\end{align}
 and if we reduce on a direction longitudinal to the M5-branes but transverse to the M2-branes we find a bound state of D2-branes and D4-branes with the alignment
\begin{align} \label{eq: D2-D4 brane config}
\begin{array}{rrrrrrr}
D2: & 0 & 1 & 2 & & &  \\
D4:& 0 & 1 & &3 &4 & 5  \ \ . \\
\end{array}
\end{align}
We recognise the second configuration as the T-dual of the D1-D3 system studied in section \ref{sect: d1-d3} on a direction transverse to both types of branes, so we can immediately see that the limit will reduce to a two-dimensional $\sigma$-model on the moduli space of the BPS monopole equations \eqref{eq: monopole equations}. The case of interest is therefore \eqref{eq: F1-D4 brane config}, which we shall study from the perspective of the D4-branes.

As discussed in section \ref{sect: D0NC-D4}, the D4NC limit of the D4-brane worldvolume theory is five-dimensional $\cal{N}=2$ SYM, which we are interested in taking the SNC limit of. Using the fundamental string's supergravity solution
\begin{subequations}
\begin{align}
    ds^2 &= H_s^{-1} \brac{-dt^2 + dX^2} + dx^i dx^i + dY^M dY^M \ , \\
    B_2 &= H_s^{-1} dt \wedge d X \ , \\
    e^{\Phi} &= H_s^{-1/2} \ ,
\end{align}
\end{subequations}
we can take the SNC limit of flat spacetime by setting $H_s=c^{-2}$. This motivates the rescaling
\begin{subequations}
\begin{align}
    \hat{t} &= c t \ , \\
    \hat{x}^i &= x^i \ ,
\end{align}
\end{subequations}
and
\begin{subequations}
\begin{align}
    \hat{X}(\hat{t}, \hat{x}) &= c X(t,x) \ , \\
    \hat{Y}^M(\hat{t}, \hat{x}) &= Y^M(t,x) \ , \\
    \hat{A}_t(\hat{t}, \hat{x}) &= c^{-1} A_t(t,x) \ , \\
    \hat{A}_i(\hat{t}, \hat{x}) &= A_i(t,x) \ ,
\end{align}
\end{subequations}
of the coordinates and fields in the field theory. We must also scale the coupling to
\begin{equation}
    \hat{g}_{YM}^2 = c g_{YM}^2 \ ,
\end{equation}
and couple the system to the flat B-field
\begin{equation}
    B_2 = c^2 dt \wedge dX \ .
\end{equation}
The effect of the B-field is to shift the field strength $F\to F-{}^*B$. In this case this simply corresponds to shifting the timelike component of the gauge field by a divergent term,
\begin{equation}
    A_t(t,x) \to \Tilde{A}_t(t,x) = A_t(t,x) + c^2 X(t,x) \ .
\end{equation}
As $X$ is in the adjoint representation $\Tilde{A}=(\Tilde{A}_t, A_i)$ is still a well-defined $SU(N)$ connection. For finite $c$ this can be undone by a field redefinition, but with this parameterisation all divergent terms in the action cancel and we can therefore take the $c\to\infty$ limit. This shift is similar to one required in the M2-brane case \cite{Lambert:2019nti, Lambert:2024uue}, and it seems likely that we can also interpret it there as arising from the coupling of the background three-form field to the M2-brane worldvolume. Taking the limit leaves us with the five-dimensional Galilean Yang-Mills theory
\begin{align} \nonumber
    S_{GYM} = \inv{2g_{YM}^2} \tr \int dt d^4x \bigg(&
    D_t X D_t X - 2 D_i X F_{0i} - \inv{2} F_{ij} F_{ij} - 2i D_t Y^M [X, Y^M] \\
    &- D_i Y^M D_i Y^M + \inv{2} [Y^M, Y^N]^2 \bigg) \ ;
\end{align}
this also follows from T-duality combined with the SNC limit of the D3-brane's worldvolume theory \cite{Fontanella:2024rvn, Lambert:2024yjk}.

To better understand the theory we should examine its classical minimum energy solutions. A quick computation shows us that the energy of a configuration is
\begin{align} \label{eq: D4-F1 energy}
    E = \inv{2 g_{YM}^2} \tr \int d^4x \bigg(
    D_t X D_t X + \inv{2} F_{ij} F_{ij} + D_i Y^M D_i Y^M - \inv{2} [Y^M, Y^N]^2 \bigg) \ .
\end{align}
To simplify the discussion let us consider the case $Y^M=0$; as each term in \eqref{eq: D4-F1 energy} is positive-definite such configurations minimise the energy. To set the first term to zero we must look for static configurations, and the obvious choice is to set $\partial_t X = A_0 = 0$; it will also be necessary to set $\partial_t A_i = 0$. The second term is minimised when the instanton equations \eqref{eq: instanton equations} are obeyed, and a static $k$-instanton configuration has the energy
\begin{equation}
    E_k = \frac{8 \pi^2 k}{g_{YM}^2} \ .
\end{equation}
All equations of motion for the system are satisfied if $X$ obeys the Gauss law constraint
\begin{equation}
    D_i D_i X = 0 \ ,
\end{equation}
which has a unique non-singular solution that is non-trivial when $X$ is given a non-zero asymptotic value. This sources an electric field for the connection $\Tilde{A}$, and so we recognise the system as describing low-energy fluctuations around the dyonic instantons \cite{Lambert:1999ua} of the relativistic theory. From the persepective of the six-dimensional theory these are the Kaluza-Klein modes of the compact direction.

\subsection{The Dual Gravitational Picture}

As with the string theory examples, we expect to be able to take two different decoupling limits of brane configurations in eleven-dimensional supergravity to obtain the bulk duals of the field theories described above. In analogy with the intersecting D-brane solution \eqref{eq: intersecting d-brane solution}, we shall generate the M-theory decoupling limits from the intersecting M2-M5 solution
\begin{subequations} \label{eq: intersecting M2-M5 solution}
\begin{align}
    \hat{G} & = H^{-\frac23}_2H^{-\frac13}_5	\eta_{\mu\nu}d\sigma^\mu d\sigma^\nu +H^{\frac13}_2  H^{-\frac13}_5dx^i d x^i + H^{-\frac23}_2H^{\frac23}_5 d X  dX + H^{\frac13}_2H^{\frac23}_5 dY^M d Y^M \ , \\
    \hat{C}_{3} & = H_2^{-1} d\sigma^0\wedge d\sigma^1\wedge  dX \ , \\
    \hat{C}_{6} & = H_5^{-1} d\sigma^0\wedge d\sigma^1 \wedge dx^2 \wedge \ldots \wedge dx^5 \ .
\end{align}
\end{subequations}

\subsubsection{The M2NC Limit of the M5-Brane}

The M2NC limit of eleven-dimensional supergravity \cite{Blair:2021waq} is found using the field redefinition
\begin{subequations}
\begin{align}
    \hat{G}_{\mn} &= c^2 \cal{T}_{\mn} + c^{-1} \cal{E}_{\mn} \ , \\
    \hat{C}_3 &= c^3 \tau^0 \wedge \tau^1 \wedge \tau^2 + C_3 \ , \\
    \hat{C}_6 &= C_6 \ ,
\end{align}
\end{subequations}
before taking the $c\to\infty$ limit. A solution with this form is obtained by taking $H_2=c^{-3}$ in \eqref{eq: intersecting M2-M5 solution}, resulting in
\begin{subequations} \label{eq: m2nc m5 solution}
\begin{align}
    \hat{G} &= c^2 \left[
    - H_5^{-1/3} \eta_{\mn} d \sigma^{\mu} d\sigma^{\nu} + H_5^{2/3} dX^2 \right] + c^{-1} \left[ H_5^{-1/3} dx^i dx^i + H_5^{2/3} dY^M dY^M \right] \ , \\
    \hat{C}_3 &= c^3 d\sigma^0 \wedge d\sigma^1 \wedge dX \ , \\
    \hat{C}_6 &= H_5^{-1} d\sigma^0 \wedge d\sigma^1 \wedge dx^2 \wedge \ldots \wedge dx^5 \ ,
\end{align}
\end{subequations}
where we smear the M5-branes along $X$ so that the harmonic function is 
 \begin{align}
    H_5 = 	 1 + \frac{R^2}{ |Y|^2}	\ .
 \end{align}
As for the D1-D3 configuration considered in section \ref{sect: d1-d3 gravity} we take $X$ to parameterise an interval. The M2NC limit possesses the Newton-Cartan-Weyl symmetry
\begin{subequations}
\begin{align}
    \cal{T}_{\mn} &\to e^{2\omega} \cal{T}_{\mn} \ , \\
    \cal{E}_{\mn} &\to e^{-\omega} \cal{E}_{\mn} \ ,
\end{align}
\end{subequations}
which we can use after taking the near-horizon limit to put the resulting geometry in the more transparent form
\begin{align} \label{eq: rescaled m2-m5 metric}
    \hat{G}' & = c^2\left[	\left(\frac{|Y|}{R}\right)^2 \eta_{\mu\nu}d\sigma^\mu d\sigma^\nu +  	dXdX\right]  + c^{-1} \left[  d x^i dx^i + \left(\frac{R}{|Y|}\right)^2	dY^M d Y^M \right] \ .
\end{align}
The most interesting set of symmetries of the solution are the infinitesimal transformations
\begin{subequations}
\begin{align}
    \sigma^{\pm} &\to \sigma^{\prime \, \pm} = \sigma^{\pm} + f^{\pm}(\sigma^{\pm}) \ , \\
    Y^M &\to Y^{\prime \, M} = \brac{1 - \partial_{\pm} f^{\pm} }Y^M \ ,
\end{align}
\end{subequations}
which leaves \eqref{eq: rescaled m2-m5 metric} invariant for any functions $f^{\pm}(\sigma^{\pm})$. We therefore find an enhancement of the relativistic conformal symmetry to two copies of the infinite-dimensional Virasoro algebra. Other than this, we as usual have the Euclidean group of transformations acting on $x^i$ and $SO(4)$ rotations of $Y^M$, giving the symmetry algebra
\begin{equation}
    \frak{g} = \frak{vir}_L \oplus \frak{vir}_R \oplus \frak{iso}(4) \oplus \frak{so}(4) \ .
\end{equation}
Aside from the translational part of the $\frak{iso}(4)$ factor, this exactly matches that of the field theory.
 
If we use the ansatz \eqref{eq: 11d reduction}, combined with the reduction
\begin{subequations}
\begin{align}
    C_3 &\to C_3 + d\sigma^1 \wedge B_2 \ , \\
    C_6 &\to B_6 - d\sigma^1 \wedge C_5 \ ,
\end{align} 
\end{subequations}
of the form fields, to compactify \eqref{eq: m2nc m5 solution} along $\sigma^1$ we find the ten-dimensional fields
\begin{subequations}
\begin{align}
    \hat{g} &= c^3 \left[ - \brac{\frac{|Y|}{R}}^3 dt^2 + \frac{|Y|}{R} dX^2 \right] + \frac{|Y|}{R} dx^i dx^i + \frac{R}{|Y|} dY^M dY^M \ , \\
    e^{\hat{\varphi}} &= \brac{\frac{c |Y|}{R}}^{3/2} \ , \\
    B_2 &= - c^3 dt \wedge d X  \ , \\
    C_5 &= \brac{\frac{|Y|}{R}}^2 dt \wedge dx^2 \wedge \ldots \wedge dx^5 \ .
\end{align}
\end{subequations}
Upon rescaling $c$ to $c^{2/3}$ and performing a Newton-Cartan-Weyl transformation we arrive at
\begin{subequations}
\begin{align}
    \hat{g} &= c^2 \left[ - H_4^{-1/2} dt^2 + H_4^{1/2} dX^2  \right] + H_4^{-1/2} dx^i dx^i + H_4^{1/2} dY^M dY^M  \ , \\
    e^{\hat{\varphi}} &= c H_4^{1/4} \ , \\
    B_2 &= - c^2 dt \wedge dX \ , \\
    C_5 &= H_4^{-1} dt \wedge dx^2 \wedge \ldots \wedge dx^5 \ ,
\end{align}
\end{subequations}
with
\begin{equation}
    H_4 = \frac{R^2}{|Y|^2} \ ,
\end{equation}
which we recognise as the near-horizon limit of the D4-brane's SNC limit. This is dual to the five-dimensional Galilean Yang-Mills theory obtained by the same procedure in section \ref{sect: MNC of (2,0)}, and can be seen to have the same symmetries.

\subsubsection{The M5NC Limit of the M2-Brane}

Finally, we consider the reciprocal limit. The M5NC limit of eleven-dimensional supergravity takes the fields to be of the form
\begin{subequations}
\begin{align}
    \hat{G}_{\mn} &= c^2 \cal{T}_{\mn} + c^{-4} \cal{E}_{\mn} \ , \\
    \hat{C}_3 &= C_3 \ , \\
    \hat{C}_6 &= c^6 \tau^0 \wedge \ldots \wedge \tau^5 \ .
\end{align}
\end{subequations}
We can generate this from the solution \eqref{eq: M2-M5 brane config} by taking $H_5=c^{-6}$, giving
\begin{subequations}
\begin{align}
    \hat{G} &= c^2 \left[ 
    - H_2^{-2/3} \eta_{\mn} d\sigma^{\mu} d\sigma^{\nu} + H_2^{1/3} dx^i dx^i \right] + c^{-4} \left[ H_2^{-2/3} dX^2 + H_2^{1/3} dY^M dY^M \right] \ , \\
    \hat{C}_3 &= H_2^{-1} d\sigma^0 \wedge d\sigma^1 \wedge dX \ , \\
    \hat{C}_6 &= c^6 d\sigma^0 \wedge d\sigma^1 \wedge dx^2 \wedge \ldots dx^5 \ ,
\end{align}
\end{subequations}
where we consider the smeared M2-brane harmonic function
\begin{align}
    H_2 & = 1 + \frac{R^2}{|Y|^2} \ .
\end{align}	
To match this with both the field theory's geometry and the solution of the previous section we take $X$ to be the coordinate of an interval. After taking the near-horizon limit and using a Newton-Cartan-Weyl transformation the metric takes the form
\begin{align}
    \hat{G}' & = c^2\left[\left(\frac{|Y|}{R}\right)^2	 \eta_{\mn} d\sigma^{\mu} d\sigma^{\nu} +  dx^i  d x^i \right]  + c^{-4} \left[ d XdX+ \left(\frac{R}{|Y|} \right)^2 dY^M d Y^M \right] \ .
\end{align}
This is identical to the smeared M2NC brane case after interchanging $X$ with $x^i$, and so we find the same set of symmetries in both cases.

\section{Conclusion} \label{sect: conclusion} 
 
 In this paper we have studied various non-Lorentzian limits of intersecting brane configurations: D0-D4, D1-D3 and M2-M5. In each case we considered the limits where one set of branes imposes an associated Newton-Cartan (NC) limit on the other. In previous work we have seen that this leads to non-Lorentzian field theories whose dynamics describes the intersection of the branes, typically in the form of a sigma-model on a BPS moduli space. It follows that we should therefore find the same dynamics whether we take an NC limit arising from the first type of brane acting on the field theory of the second or vice-versa. In this paper we have explicitly demonstrated this. Furthermore we   considered NC limits of the dual near horizon geometries and showed that the symmetries match those of the field theories. Thus we find four different descriptions of the same physical system: two from non-Lorentzian field theories and two from their gravitational duals. 

The dynamics associated to D0-D4 and D1-D3 brane intersections are relatively well-known. In particular they are described by sigma-models on instanton moduli space and monopole moduli space respectively. Furthermore quantum mechanics on instanton moduli space is known to have an $SO(2,1)$ conformal symmetry and we have proposed two gravitational duals. For monopole moduli space the dynamics takes place on a Coulomb branch where, although there is a scale symmetry of the non-Lorentzian action,  it is spontaneously broken and this is reflected in the gravitational duals by a running dilaton. We also presented a new example arising from M2-branes ending on M5-branes. This leads to a two-dimensional sigma model on the moduli space of solutions to the Basu-Harvey equation. The non-Lorentzian field theory admits separate left and right copies of the Virasoro algebra which naturally descends to the sigma-model. We constructed the gravitational duals in Newton-Cartan geometry and showed that they also admit the infinite-dimensional symmetry algebras. The associated dynamics represent excitations of the so-called self-dual string solution and should be interpreted as  arising from the enigmatic six-dimensional $\cal{N}=(2,0)$ theory. We also presented novel NC limits of the Bosonic sectors of type IIA and type IIB supergravity.

We believe that the examples presented here can be generalised and extended. It would be useful to see how the limits discussed here should be applied at a more fundamental level in the bulk duals, where in general the description should be in terms of Matrix $p$-Theory \cite{Blair:2023noj, Blair:2024aqz} in the $p$NC background. It is also of interest to see how these limits act on the closed string worldsheet CFT that defines the perturbative expansion of the relativistic theory. These theories are known to be integrable and one might hope that this remains true once the limit is taken. We also hope that non-Lorentzian limits can be used to gain new insight into M-theory.

\section*{Acknowledgements}
 
 N.L. is supported in part  by the STFC consolidated grant ST/X000753/1. J.S. is supported by the STFC studentship ST/W507556/1.

\appendix

\section{The Moduli Space Description} \label{sect: appendix moduli space}

\subsection{The D0-D4 System} \label{sect: appendix moduli space d0-d4}

In order to show the equivalence of the two limit orders for the D0-D4 system studied in section \ref{sect: d0-d4} it will be helpful to review the reduction of the action \eqref{eq: D0NC action} to its moduli space description. There are many fantastic introductions to this topic, see for instance \cite{Tong:2005un, Dorey:2002ik, Vandoren:2008xg}. Let us set $Y^M=0$ momentarily. Imposing the equation of motion for $G_{ij}$ forces our $SU(N)$ gauge field to obey the instanton equations
\begin{equation}
    F = * F ;
\end{equation}
let us denote the general solution to these for instanton number $k$ as $A_i(x;m)$, where $m^{I}$ are moduli that parameterise different solutions. They form a set of coordinates for the $4kN$-dimensional moduli space $\cal{M}_{k,N}$. As the instanton equations do not involve time these remain solutions if we give the moduli time-dependence. It will be convenient to fix the gauge we work in such that timelike component of the gauge field is
\begin{equation} \label{eq: moduli space A0 expression}
    A_0 = \dot{m}^{I} \omega_{I} \ ,
\end{equation}
with $\omega_{I}$ the components of an $\frak{su}(N)$-valued one-form on $\cal{M}_{k,N}$. The electric field in this gauge is
\begin{equation}
    F_{0i} = \dot{m}^{I} \brac{\partial_{I} A_i - D_i \omega_{I}} \equiv \dot{m}^{I} \delta_{I} A_i \ .
\end{equation}
The fields $\delta_{I} A_i$ satisfy the linearised form of the instanton equations around another solution and are known as zero-modes. This choice of gauge is valid as long as Gauss's law
\begin{equation} \label{eq: gauss law}
    D_i \delta_{I} A_i = 0
\end{equation}
is obeyed, which fixes $\omega_{I}$. This ensures that zero-modes are orthogonal to gauge-transformations which vanish at the asymptotic spatial boundary. However, the large gauge transformations which are non-vanishing at the boundary are not projected out in this manner. As the instanton equations are gauge-invariant they automatically solve their linearised form and must therefore must be thought of as genuine zero-modes with associated moduli. 

With this parameterisation, the action \eqref{eq: D0NC action} can be written as
\begin{subequations} \label{eq: moduli space kinetic term reduction}
\begin{align}
    S_{D0NC} = \inv{2} \int dt \, g_{IJ} \dot{m}^I \dot{m}^J \ , \\ 
    g_{IJ} = \inv{g_{YM}^2} \tr \int d^4x \, \delta_{I} A_i \delta_{J} A_ i \ .
\end{align}
\end{subequations}
It can be shown that $g_{IJ}$ defines a hyper-K\"ahler metric on $\cal{M}_{k,N}$, so the content of the action reduces to the quantum mechanics of a free non-relativistic particle on this manifold.

Let us now allow non-zero values for the scalar fields $Y^M$. These obey \eqref{eq: Y spatial laplacian}, meaning their term in the action can equivalently be written as the boundary term
\begin{equation}
    S_{Y} = -\inv{2 g_{YM}^2} \tr \int dt \int_{S^3_{\infty}} d S_i \, \Big( r^2 x^i Y^M D_i Y^M \Big)\Big\rvert_{r\to\infty} \ .
\end{equation}
We must therefore allow $Y^M$ to be non-vanishing at the boundary in order to have a non-zero potential. However, there is a different way to think of the potential that makes its relation to the moduli space clear using an adaptation of the argument in \cite{Tong:1999mg}. The equation the scalars satisfy can be seen to be a special case of Gauss's law \eqref{eq: gauss law} for the case of a gauge transformation $\delta A_i = D_i \Omega$. As $Y^M$ must be non-vanishing at the boundary we can expand it in a basis of gauge zero-modes, giving
\begin{equation} \label{eq: zero mode expansion}
    D_i Y^M = \lambda^{M,r} K_r^I \delta_I A_i \equiv K_{\lambda^M}^I \delta_I A_i \ ,
\end{equation}
where $r$ labels the gauge zero-modes, $\lambda^{M,r}$ are a set of parameters, and $K_r^I$ the components of a set of vectors on $\cal{M}_{k,N}$. Substituting \eqref{eq: zero mode expansion} into the action yields
\begin{equation} \label{eq: D0NC potential}
    S_Y = - \inv{2} \int dt \, g_{IJ} K^I_{\lambda^M} K^J_{\lambda^M} \ ,
\end{equation}
and so the potential is the magnitude of the linear combination of vectors $K_{\lambda^M}$. As the large gauge transformations are a non-trivial symmetry of the theory they must also descend to symmetries of the moduli space; in fact, these symmetries are generated by the vectors $K^I_r$, which one can show are actually tri-holomorphic Killing vectors.

We can now repeat the analysis for the action \eqref{eq: D4NC action 2}. As before, let us set $\Tilde{Y}^M$ to zero initially. The equations of motion for $G^i$ impose the ADHM equations
\begin{equation}
    \calz_a^{\alpha} \sigma^i_{\alpha\beta} \calz^{\dag \, \beta a} - i \bar{\eta}^i_{AB} [X^A, X^B] = 0\ ,
\end{equation}
on the dynamical fields. The most general solutions to these are
\begin{subequations}
\begin{align}
    X^A = X^A(m(t)) \ , \\
    \calz^{\alpha}_a = \calz^{\alpha}_a(m(t)) \ ,
\end{align}
\end{subequations}
where $m^I$ again denote the moduli and we have allowed for time-dependence. We can completely gauge away the worldline gauge-field, so
the kinetic terms in the action become
\begin{subequations}
\begin{align}
    S_{D4NC} &= \inv{2} \int dt \, \Tilde{g}_{IJ} \dot{m}^I \dot{m}^J \ , \\
    \Tilde{g}_{IJ} &= \inv{g_{YM}^2} \tr \brac{
    \partial_I X^A \partial_J X^A + 2 \partial_{(I} \calz^{\alpha}_a \partial_{J)} \calz^{\dag\, \alpha a}} \ .
\end{align}
\end{subequations}
As mentioned above the metric defined here is isomorphic to the instanton case, so the kinetic terms here are just those in \eqref{eq: moduli space kinetic term reduction}. 

We now consider the case where $\Tilde{Y}^M$ is non-zero. In order to see that this gives the same potential as the previous theory, we note that as $\calz^{\alpha}_a$ is in the antifundamental of $SU(N)$ the combination $\calz^{\alpha}_b \tensor{(\Tilde{Y}^M)}{^b_a}$ is the action of an $\frak{su}(N)$-generator on the field. Since these transformations are isometries of the moduli space that preserve the complex structures they can be expanded in the same basis of Killing vectors that generated the $SU(N)$ transformations before, meaning we find
\begin{equation}
    \calz^{\alpha}_b \tensor{(\Tilde{Y}^M)}{^b_a} = \lambda^{M,r} K^I_{r} \partial_I \calz^{\alpha}_a \equiv K_{\lambda^M}^I \partial_I \calz^{\alpha}_a \ .
\end{equation}
Since $X^A$ is invariant under the $SU(N)$ transformations its zero-modes are orthogonal to these vectors, so
\begin{equation}
    K^I_r \partial_I X^A = 0 \ .
\end{equation}
The potential terms can then be rewritten as
\begin{equation}
    S_{\Tilde{Y}} = - \inv{2} \int dt \, \Tilde{g}_{IJ} K^I_{\lambda^M} K^J_{\lambda^M} \ ,
\end{equation}
which we see is identical to \eqref{eq: D0NC potential} due to the fact the metrics are isomorphic.

\subsection{The D1-D3 System} \label{sect: appendix moduli space d1-d3}

We will now perform the same analysis for the D1-D3 system discussed in section \ref{sect: d1-d3}, starting with the monopole theory with action \eqref{eq: D1NC action}. As this is incredibly similar to the procedure for the instanton theory we will be somewhat brief; for a comphrensive review see \cite{Weinberg:2006rq} and the references therein. 

As in the previous cases we will start by setting the transverse scalars $Y^M$ to zero in \eqref{eq: D1NC action}. The Hubbard-Stratonovich field imposes the BPS monopole equations \eqref{eq: monopole equations}
\begin{equation}
    F_{ij} = \epsilon_{ijk} D_k X \ .
\end{equation}
We will denote the general solution to this in the $k$-monopole sector by $A_i(x;m(t))$ and $X(x,m(t))$, where as for the instanton theory $m^{I}$ are coordinates that parameterise the moduli space $\Sigma_{k,N}$ and all time-dependence enters through the moduli. The timelike component of the gauge field is again given by \eqref{eq: moduli space A0 expression}, where now $\omega_I$ is constrained by Gauss's law to satisfy
\begin{equation} \label{eq: monopole gauss law}
    D_i \delta_{I} A_i - i [X,\delta_{I} X] = 0 \ ,
\end{equation}
with
\begin{subequations}
\begin{align}
    \delta_{I} A_i = \partial_{I} A_i - D_i \omega_I \ , \\
    \delta_I X = \partial_I X - i [\omega_I, X ] \ .
\end{align}
\end{subequations}
A metric on the moduli space is then given by
\begin{equation} \label{eq: monopole metric}
    G_{IJ} = \inv{g_{YM}^2} \tr \int d^3x \brac{
    \delta_I A_i \delta_J A_i + \delta_I X \delta_J X} \ ,
\end{equation}
and the action \eqref{eq: D1NC action} reduces to
\begin{equation} \label{eq: D1NC moduli space reduction}
    S_{D1NC} = \inv{2} \int dt \, G_{IJ} \dot{m}^I \dot{m}^J \ .
\end{equation}

Allowing for non-zero values of $Y^M$ proceeds in the same way as before. The scalars obey the equation of motion
\begin{equation}
    D_i D_i Y^M - [X,[X,Y^M]] = 0 \ ,
\end{equation}
with a non-trivial boundary value of $Y^M$, which using the argument of \cite{Tong:1999mg} we see is identical to \eqref{eq: monopole gauss law} for the case of a large gauge transformation,
\begin{subequations}
\begin{align}
    \delta_M A_i = D_i Y^M \ , \\
    \delta_M X = i [Y^M, X] \ .
\end{align}
\end{subequations}
Unlike in the instanton case we have the added subtlety that the VEV of $X$ spontaneously breaks the global gauge symmetry to a subgroup. For simplicity we will assume the case of maximal symmetry breaking ({\it i.e.} all the D3-branes are seperated, so none of the eigenvalues of $X$ coincide), in which case the preserved subgroup which are global is $U(1)^{N-1}$. As the global gauge transformations are both zero-modes of the BPS monopole equations and symmetries of the system, we can expand $Y^M$ in a basis of the associated tri-holomorphic Killing vectors on $\Sigma_{k,N}$,
\begin{subequations}
\begin{align}
    D_i Y^M &= \lambda^{M,r} K_r^I \delta_I A_i \equiv K_{\lambda^M}^I \delta_I A_i \ , \\
    i [Y^M,X] &= \lambda^{M,r} K_r^I \delta_I X \equiv K_{\lambda^M}^I \delta_I X \ ,
\end{align}
\end{subequations}
as in the instanton case. Repeating the steps outlined in the previous section, we see that the terms involving $Y^M$ in \eqref{eq: D1NC action} become
\begin{equation}
    S_{Y} = - \inv{2} \int dt \, G_{IJ} K^I_{Y^M} K^J_{Y^M} \ .
\end{equation}

Let us now do the same for the D3NC theory \eqref{eq: D1-D3NC action}, starting with the case $Y^M = 0$ as usual. The constraint equations we must impose are now Nahm's equations \eqref{eq: Nahm equations},
\begin{equation}
    D_x X^i = \frac{i}{2} \epsilon_{ijk} [X^j, X^k] \ ,
\end{equation}
on the interval $x\in[-L,L]$. We will focus here on the case of $k$$SU(2)$ monopoles; for these, we must solve \eqref{eq: Nahm equations} subject to the condition that there is a simple pole in $X^i$ at each endpoint, {\it i.e.}
\begin{equation}
    \lim_{x\to L}(x-L) X^i = t^i\ ,
\end{equation}
and similarly for $x=-L$, whose residues $t^i$ define the $k$-dimensional irreducible representation of $\frak{su}(2)$. We can then proceed as above; solving \eqref{eq: Nahm equations} and defining the zero-modes
\begin{subequations}
\begin{align}
    \delta_I A_x &= \partial_I A_x - D_x \omega_I \ , \\
    \delta_I X^i &= \partial_I X^i - i [\omega_I, X] \ ,
\end{align}
\end{subequations}
with $A_0 = \omega_I \dot{m}^I$ determined by the Gauss law constraint
\begin{equation} \label{eq: d3nc gauss law}
    D_x \delta_I A_x - i [X^i , \delta_I X^i] = 0 \ ,
\end{equation}
we can use the hyper-K\"ahler isometry \cite{Nakajima:1990} between the moduli spaces of the $SU(2)$ BPS monopole equations in the $k$-monpole sector and $U(k)$ Nahm's equations with metrics \eqref{eq: monopole metric} and 
\begin{equation} \label{eq: Nahm metric}
    \Tilde{G}_{IJ} = \inv{g_{YM}^2} \tr \int_{-L}^L dx \brac{
    \delta_I A_x \delta_J A_x + \delta_I X^i \delta_J X^i } \ ,
\end{equation}
to show that the D3NC theory reduction
\begin{equation}
    S_{D3NC} = \inv{2} \int dt \, \Tilde{G}_{IJ} \dot{m}^I \dot{m}^j \ ,
\end{equation}
is equal to \eqref{eq: D1NC moduli space reduction}. One may initially worry that the poles in $X^i$ render \eqref{eq: Nahm metric} divergent; however, as the residues of the poles are moduli-independent the zero-modes $\delta_I X^i$ are regular and the metric is well-defined. The extension to the general $SU(N)$ theory is conceptually clear but technically convoluted. For the case of maximal symmetry breaking the interval is broken into $N-1$ subintervals on which Nahm's equations are solved, with a generically different value of $k$ in each subinterval. Boundary conditions are then imposed at the points where the intervals meet (in the string theory context these are the D3-brane positions), with different conditions arising depending on the difference in $k$ between the two sides. One would imagine that a similar isometry can be found between the moduli space of these solutions and solutions to the $SU(N)$ monopole equations, though to our knowledge a proof of this does not exist.

The potential obtained by turning on the scalar fields $Y^M$ is essentially analogous to the D1NC case. The equation the fields satisfy is now
\begin{equation}
    D_x D_x Y^M - [X^i, [X^i, Y^M]] = 0 \ ,
\end{equation}
which is again the Gauss law constraint \eqref{eq: d3nc gauss law} for large gauge transformations, and so a similar construction to that above can be used to show that the potential is the norm of a tri-holomorphic Killing vector on the moduli space associated with these symmetries. However, in order for this to be true we must also show that the large gauge symmetry groups match on both sides. For the $SU(2)$ monopoles this is just $U(1)$. To see this in the D3NC theory, we note that as all fields are in the adjoint of $U(k)$ the $U(1)$ part of $A_x$ drops out Nahm's equations. We can therefore completely remove the non-Abelian part of $A_x$ from the theory by gauge-fixing it to zero with a $\frak{su}(k)$-valued gauge transformation, with the non-Abelian part of $A_0$ fixed by \eqref{eq: d3nc gauss law}. This leaves behind a decoupled two-dimensional Abelian gauge theory, whose large gauge transformations form the symmetry group in question. The extension to the general case is then clear- this procedure can be replicated in each of the subintervals, with the large gauge transformations now those that don't vanish at the D3-brane positions. This leads to a $U(1)^{N-1}$ symmetry group, in agreement with the D1NC symmetry group.

\section{Limits of Supergravity Actions and Symmetries}

\subsection{The D0NC Limit of Type IIA Supergravity} \label{sect: iia d0nc action}

To take the limit we use the decomposition \cite{Blair:2023noj} \eqref{eq: D0NC decomp} for the Bosonic fields of type IIA supergravity, where we shall work in the string frame. The fields $\tau$ and $\cE$ that arise from the metric satisfy the relations
\begin{subequations}
\begin{align}
    \tau_{\mu} \cE^{\mn} &= 0 \ , \\
    \tau^{\mu} \cE_{\mn} &= 0 \ , \\
    -\tau^{\mu} \tau_{\nu} + E^{\mu\rho} E_{\rho\nu} &= \delta^{\mu}_{\nu} \ , \\
    \tau^{\mu} \tau_{\mu} &= -1 \ .
\end{align}
\end{subequations}
The decomposition of the metric into a clock one-form $\tau_{\mu}$ and spatial cometric $\cE^{\mn}$ is familiar from Newton-Cartan limits of GR\footnote{See \cite{Hartong:2022lsy} and the references therein for a thorough review of this area.}. The Bosonic action of type IIA supergravity in the string frame is
\begin{align} \nonumber
    \hat{S}_{IIA} = \inv{4 \kappa^2} \bigintsss \Bigg( d^{10}x \sqrt{-\hat{G}} \bigg[ &2e^{-2\hat{\varphi}} \bigg( \hat{R}(\hat{G}) + 4 \partial_{\mu} \hat{\varphi} \partial^{\mu} \hat{\varphi} - \inv{12} \hat{H}_{\mu\nu\rho} \hat{H}^{\mu\nu\rho} \bigg) 
    - \inv{2} \hat{F}_{\mu\nu} \hat{F}^{\mn} \\
    &- \inv{4!} \hat{\Tilde{F}}_{\mn\rho\lambda} \hat{\Tilde{F}}^{\mn\rho\lambda} \bigg] -  \hat{B}_2 \wedge \hat{F}_4 \wedge \hat{F}_4 \Bigg) \ ,
\end{align}
where $\hat{\Tilde{F}}_4$ is given by
\begin{equation}
    \hat{\Tilde{F}}_4 = \hat{F}_4 - \hat{C}_1 \wedge \hat{H}_3 \ .
\end{equation}
Inserting the decomposition \eqref{eq: D0NC decomp} and taking the $c\to \infty$ limit gives
\begin{align} \nonumber
    S_{D0NC} = \inv{4\kappa^2} \bigintsss \Bigg( d^{10}x \,\Omega \bigg[ &
    2 e^{-2\varphi} \cE^{\mn} \check{R}_{\mn} - 16 e^{-2\varphi} \cE^{\mn} \partial_{\mu} \varphi \partial_{\nu} \varphi - e^{-\varphi} \cE^{\mn} \cE^{\rho\sigma} F_{\mu\rho} T_{\nu\sigma} \\ \nonumber
    &+ \inv{2} e^{-2\varphi} \tau^{\mu_1} \tau^{\nu_1} E^{\mu_2\nu_2} \cE^{\mu_3\nu_3} H_{\mu_1 \mu_2 \mu_3} H_{\nu_1 \nu_2 \nu_3} \\ \nonumber
    & + e^{-2\varphi} \cE^{\mn} \brac{\tau^{\rho} T_{\rho \mu} - 5 \partial_{\mu} \varphi} \brac{\tau^{\sigma} T_{\sigma \nu} - 5 \partial_{\nu} \varphi}  \\ \nonumber
    &- \inv{4!} \cE^{\mu_1\nu_1} \cE^{\mu_2\nu_2} \cE^{\mu_3\nu_3} \cE^{\mu_4\nu_4} \Tilde{F}_{\mu_1 \ldots \mu_4} \Tilde{F}_{\nu_1 \ldots \nu_4} \\ \nonumber
    &+ \inv{3} e^{-\varphi} \tau^{\mu_1} \cE^{\mu_2\nu_2} \cE^{\mu_3\nu_3} \cE^{\mu_4\nu_4} \Tilde{F}_{\mu_1 \ldots \mu_4} H_{ \nu_2 \nu_3 \nu_4} \bigg] \\ \label{eq: D0NC gravity action}
    &-  B_2 \wedge F_4 \wedge F_4 \Bigg) \ ,
\end{align}
where $\check{R}_{\mn}$ is the Ricci tensor of the generically torsional connection \cite{Hansen:2020pqs}
\begin{equation}
    \check{\Gamma}_{\mn}^{\rho} = - \tau^{\rho} \partial_{\mu} \tau_{\nu} + \inv{2} \cE^{\rho\sigma} \brac{\partial_{\mu} \cE_{\nu\sigma} + \partial_{\nu} \cE_{\mu\sigma} - \partial_{\sigma} \cE_{\mn} } \ ,
\end{equation}
the tensor $T_{\mn}$ is the torsion of the connection,
\begin{equation}
    T_{\mn} = 2 \partial_{[\mu} \tau_{\nu]} \ ,
\end{equation}
and $\Omega$ is defined to be
\begin{equation}
    \Omega = \sqrt{-\det\brac{-\tau_{\mu} \tau_{\nu} + \cE_{\mn}}} \ .
\end{equation}
Remarkably, all the divergent terms that arise in the rescaling cancel, leaving a finite action.

The action \eqref{eq: D0NC gravity action} has a number of gauge symmetries. We retain the diffeomorphism invariance and $p$-form gauge-symmetry of the original relativistic action. In addition, there are also new symmetries that emerge once the limit has been taken. The first of these are local Galilean boosts, parameterised by a one-form $\lambda$ that satisfies
\begin{equation}
    \tau^{\mu} \lambda_{\mu} = 0 \ .
\end{equation}
We see that the infinitesimal transformations
\begin{subequations}
\begin{align}
    \delta \tau_{\mu} &= c^{-4} \lambda_{\mu} \ , \\
    \delta \tau^{\mu} &= \cE^{\mn} \lambda_{\nu} \ , \\
    \delta \cE_{\mn} &= 2 \lambda_{(\mu} \tau_{\nu)} \ , \\
    \delta \cE^{\mn} &= - 2 c^{-4} \cE^{\rho (\mu} \tau^{\nu)} \lambda_{\rho} \ , \\
    \delta C_1 &= - e^{-\varphi} \lambda \ ,
\end{align}
\end{subequations}
leave the relativistic fields in \eqref{eq: D0NC decomp} invariant. They therefore automatically form a gauge symmetry of the action; as there are no divergences in the transformations they survive the D0NC limit, with the action \eqref{eq: D0NC gravity action} invariant under
\begin{subequations} \label{eq: local g-boost D0NC}
\begin{align}
    \delta \tau^{\mu} &= \cE^{\mn} \lambda_{\nu} \ , \\
    \delta \cE_{\mn} &= 2 \tau_{(\mu} \lambda_{\nu)} \ , \\
    \delta C_1 &= - e^{-\varphi} \lambda \ ,
\end{align}
\end{subequations}
as can be checked explicitly. The other emergent symmetry is the Newton-Cartan-Weyl scaling symmetry
\begin{subequations} \label{eq: NCW transformation}
\begin{align}
    \tau'_{\mu} &= e^{\omega} \tau_{\mu} \ , \\
    \cE^{\prime \, \mn} &= e^{2 \omega} \cE^{\mn} \ , \\
    \varphi' &= \varphi - 3 \omega \ ,
\end{align}
\end{subequations}
where $\omega(x)$ is an arbitrary function. When $\omega$ is constant this amounts to a rescaling of $c$, so as the limit isolates the $c$-independent part of the relativistic action it is immediately clear that the constant $\omega$ transformations are a symmetry. As before, an explicit calculation shows that the symmetry is maintained for any $\omega(x)$.

It has long been known \cite{Duval:1984cj} that the null reduction of a Lorentzian manifold gives us a Newton-Cartan geometry. In the current context, it has been proposed that the D0NC limit is equivalent to a null reduction of M-theory \cite{Blair:2023noj}. We can explicitly check this at the level of the supergravity actions. Suppose that we work with the decomposition \eqref{eq: 11d sugra null reduction field def} of the metric and three-form field, where the fields are taken to be independent of $x^+$. Unlike the case of a spacelike circle reduction, there is no dilaton in our decomposition ansatz for the metric; its appearance is unnecessary as it can always be absorbed into the choice of Newton-Cartan metric fields and does not represent an independent degree of freedom. If we choose to include it we must therefore have a scaling symmetry akin to \eqref{eq: NCW transformation} that can remove it. Taking $x^+$ to be a periodic variable with periodicity $2\pi R$, it is straightforward to plug \eqref{eq: 11d sugra null reduction field def} into the relativistic action
\begin{align}
    \hat{S}_{11d} = \inv{2\kappa_{11}^2} \bigintsss \Bigg( d^{11}x \sqrt{- \hat{g}} \bigg[\hat{R}(\hat{g}) - \inv{48} \hat{F}_{ABCD} \hat{F}^{ABCD} \bigg] - \inv{6} \hat{C}_3 \wedge \hat{F}_4 \wedge \hat{F}_4 \Bigg) \ ,
\end{align}
and integrate over $x^+$, with the outcome being that we recover the action \eqref{eq: D0NC gravity action} with $\varphi=0$ after making the identification $\kappa^2 = \kappa_{11}^2/2\pi R$.

\subsection{The D1NC Limit of Type IIB Supergravity} \label{sect: appendix d1nc sugra limit}

Let us now compute the action for the D1NC limit of type IIB supergravity using the field redefinition \eqref{eq: D1NC gravity field redef}. The action for the Bosonic sector of type IIB supergravity in the string frame is
\begin{align} \nonumber
    \hat{S}_{IIB} = \inv{4 \kappa^2} \bigintsss \Bigg( d^{10}x \sqrt{-\hat{G}} \bigg[ &2e^{-2\hat{\varphi}} \bigg( \hat{R}(\hat{G}) + 4 \partial_{\mu} \hat{\varphi} \partial^{\mu} \hat{\varphi} - \inv{12} \hat{H}_{\mu\nu\rho} \hat{H}^{\mu\nu\rho} \bigg) 
    - \hat{F}_{\mu} \hat{F}^{\mu} \\
    &- \inv{3!} \hat{\Tilde{F}}_{\mn\rho} \hat{\Tilde{F}}^{\mn\rho} - \inv{2\cdot5!} \hat{\Tilde{F}}_{\mn\rho\lambda\sigma} \hat{\Tilde{F}}^{\mn\rho\lambda\sigma} \bigg] -  \hat{C}_4 \wedge \hat{H}_3 \wedge \hat{F}_3 \Bigg) \ ,
\end{align}
where we use the definitions
\begin{subequations}
\begin{align}
    \Tilde{F}_3 &= F_3 - C_0 H_3 \ , \\ \label{eq: tilde F5 def}
    \Tilde{F}_5 &= F_5 - C_2 \wedge H_3 \ ,
\end{align}
\end{subequations}
and must impose the additional constraint
\begin{equation} \label{eq: type iib constraint}
    \hat{\Tilde{F}}_5= * \hat{\Tilde{F}}_5  \ .
\end{equation}
Note that the definition \eqref{eq: tilde F5 def} differs slightly from the conventional one, and is reached by a field redefinition of $C_4$ that only changes the action by a boundary term; we will not worry about such terms so for our purposes the two are equivalent, with the form given above more convenient to work with. As with the D0NC theory we will denote the scaled version of the measure by
\begin{equation}
    \sqrt{-\hat{G}} = c^{-6} \Omega \ ,
\end{equation}
where $\Omega$ is given by
\begin{equation}
    \Omega = \sqrt{- \det\brac{\cal{T}_{\mn} + \cal{E}_{\mn}}} \ .
\end{equation}
Inserting \eqref{eq: D1NC gravity field redef} into the relativistic action, we see that there are divergent terms
\begin{align} \nonumber
    \hat{S}_{IIB,c^4} = -\frac{c^4}{4\kappa^2} \bigintsss \Bigg(& d^{10}x \,\Omega \bigg[ \frac{2}{4!} e^{-2\varphi} \cal{E}^{\mu_1 \nu_1} \cal{E}^{\mu_2 \nu_2} \cal{E}^{\mu_3 \nu_3} H_{\mu_1\mu_2\mu_3} H_{\nu_1\nu_2\nu_3}  \\
    &+\inv{2\cdot 5!} \cal{E}^{\mu_1 \nu_1} \ldots \cal{E}^{\mu_5 \nu_5} \Tilde{F}_{\mu_1 \ldots \mu_5} \Tilde{F}_{\nu_1 \ldots \nu_5} \bigg] + e^{-\varphi} \tau^0 \wedge \tau^1 \wedge \tilde{F}_5 \wedge H_3  \Bigg) \ .
\end{align}
Up to a positive rescaling of $\varphi$ these are exactly the divergent terms found in the SNC limit of type IIB supergravity \cite{Bergshoeff:2023ogz} after interchanging $B_2$ and $C_2$ and flipping the sign of $\varphi$, which is consistent with the notion of these limits being S-dual to one another. We therefore proceed as in that paper by noting that the divergent terms can be rewritten in the positive-definite form
\begin{align}
    \hat{S}_{IIB,c^4} = - \frac{c^4}{8\cdot 3! \, \kappa^2} \bigintsss d^{10}x \, \Omega \brac{e^{-\varphi} H_{a_1 a_2 a_3} - \inv{5!} \epsilon_{a_1 a_2 a_3 b_1 \ldots b_5} \tilde{F}_{b_1 \ldots b_5}}^2 \ ,
\end{align}
where we have rewritten the forms by performing the contractions with the orthonormal basis $e_a$ for the cometric $\cal{E}^{\mn}$. As usual we can introduce a Hubbard-Stratonovich purely-spatial three-form field $\mathscr{H}_{a_1 a_2 a_3}$ to put the action into the form
\begin{align} \nonumber
    \hat{S}_{IIB,c^4} = \inv{4\kappa^2} \bigintsss d^{10}x \, \Omega \Bigg(&
    \mathscr{H}_{a_1 a_2 a_3} \brac{e^{-\varphi} H_{a_1 a_2 a_3} - \inv{5!} \epsilon_{a_1 a_2 a_3 b_1 \ldots b_5} \tilde{F}_{b_1 \ldots b_5}} \\
    &+ \frac{3}{c^4} \mathscr{H}_{a_1 a_2 a_3} \mathscr{H}_{a_1 a_2 a_3} \Bigg) \ .
\end{align}
This allows the $c\to \infty$ limit to be safely taken, with the result being the action
\begin{align} \nonumber
    S_{D1NC} = \inv{4\kappa^2} \bigintsss \Bigg( d^{10} x \, \Omega \bigg[&2 e^{-2\varphi} \check{R} +  8 e^{-2\varphi} \cal{E}^{\mn} \partial_{\mu} \varphi \partial_{\nu} \varphi
    - \inv{3!} \cal{E}^{\mu_1 \nu_1} \cal{E}^{\mu_2 \nu_2} \cal{E}^{\mu_3 \nu_3} \Tilde{F}_{\mu_1 \mu_2 \mu_3} \Tilde{F}_{\nu_1 \nu_2 \nu_3}
    \\ \nonumber
    &+ \mathscr{H}_{a_1 a_2 a_3} \brac{e^{-\varphi} H_{a_1 a_2 a_3} - \inv{5!} \epsilon_{a_1 a_2 a_3 b_1 \ldots b_5} \tilde{F}_{b_1 \ldots b_5}} \\ \nonumber
    &- e^{-\varphi} \epsilon_{AB} \eta^{BC} \tau_C^{\nu_1} E^{\mu_2 \nu_2} E^{\mu_3 \nu_3} T^A_{\mu_2 \mu_3} \tilde{F}_{\nu_1 \nu_2 \nu_3} 
    \\ \nonumber
    &- 2 e^{-2\varphi} \eta^{AC} \eta^{BD} E^{\mn} \brac{\epsilon_{E[A} 
    \tau^{\rho}_{B]} T^E_{\rho\mu} - \epsilon_{AB} \partial_{\mu} \varphi} \brac{\epsilon_{F[C} 
    \tau^{\sigma}_{D]} T^E_{\sigma\nu} - \epsilon_{CD} \partial_{\nu} \varphi} \\ \nonumber
    &- \inv{2}e^{-2\varphi} \tau^{\mu_1 \nu_1} E^{\mu_2 \nu_2} \cal{E}^{\mu_3 \nu_3} H_{\mu_1\mu_2\mu_3} H_{\nu_1\nu_2\nu_3} \\ \nonumber
    &- \inv{2\cdot 4!} \tau^{\mu_1 \nu_1} \cal{E}^{\mu_2 \nu_2} \ldots \cal{E}^{\mu_5 \nu_5} \tilde{F}_{\mu_1 \ldots \mu_5} \tilde{F}_{\nu_1 \ldots \nu_5} \\ \nonumber
    &+ \inv{2\cdot 3!} e^{-\varphi} \epsilon^{AB} \cal{E}^{\mu_1 \nu_1} \cal{E}^{\mu_2 \nu_2} \cal{E}^{\mu_3 \nu_3} \tau_A^{\nu_4} \tau_B^{\nu_5} H_{\mu_1 \mu_2 \mu_3} \tilde{F}_{\nu_1 \ldots \nu_5}
    \bigg] \\
    &  - C_2 \wedge \tilde{F}_5 \wedge H_3
    \Bigg) \ ,
\end{align}
where $\check{R}$ is defined to be the finite part of the expansion of the Einstein-Hilbert term\footnote{See \cite{Bergshoeff:2023ogz} for an explicit expression for these terms.}, and $T_{\mu\nu}^A$ is given by
\begin{equation}
    T_{\mn}^A = \partial_{\mu} \tau_{\nu}^A - \partial_{\nu} \tau_{\mu}^A \ .
\end{equation}
The Hubbard-Stratonovich field now imposes the constraint
\begin{equation} \label{eq: iib d1nc limit constraint}
    e^{-\varphi} H_{a_1 a_2 a_3} = \inv{5!} \epsilon_{a_1 a_2 a_3 b_1 \ldots b_5} \tilde{F}_{b_1 \ldots b_5} \ .
\end{equation}
Upon making the field redefinition the self-duality constraint can be written as the two equations
\begin{subequations} \label{eq: split constraint}
\begin{align} \label{eq: split constraint 1}
    \hat{\tilde{F}}_{01 a_1 a_2 a_3} &= - \frac{c^4}{5!} \epsilon_{a_1 a_2 a_3 b_1 \ldots b_5} \hat{\tilde{F}}_{b_1 \ldots b_5} \ , \\
    \hat{\tilde{F}}_{A a_1 \ldots a_4} &= \inv{4!} \epsilon_{AB} \epsilon_{a_1 \ldots a_4 b_1 \ldots b_4} \hat{\tilde{F}}^{B b_1 \ldots b_4} \ ,
\end{align}
\end{subequations}
so as 
\begin{equation}
    \hat{\tilde{F}}_5 = - c^4 e^{-\varphi} \tau^0 \wedge \tau^1 \wedge H_3 + \tilde{F}_5 \ ,
\end{equation}
we see that the divergent pieces of \eqref{eq: split constraint 1} just recover the constraint \eqref{eq: iib d1nc limit constraint} imposed by $\mathscr{H}_3$, with the rest \eqref{eq: split constraint} being identical to the original self-duality constraint.

We can determine the non-relativistic gauge symmetries of the D1NC limit in the same way as was done for the D0NC limit in section \ref{sect: iia d0nc action} and for the SNC limit in \cite{Bergshoeff:2023ogz}. The local Galilean boost symmetry that arises after taking the $c\to \infty$ limit is
\begin{subequations}
\begin{align}
    \delta \tau^{\mu}_A &= e^{\mu}_a \lambda^a_A \ , \\
    \delta e^a_{\mu} &= - \lambda^a_A \tau_{\mu}^A \ , \\
    \delta C_2 &= - e^{-\varphi} \epsilon_{AB} \lambda^A_a e^a \wedge \tau^B \ ,
\end{align}
\end{subequations}
where $\lambda^A_a$ are a set of coefficients with indices raised and lowered by $\eta_{AB}$ and $\delta_{ab}$, and we again have an emergent Newton-Cartan-Weyl symmetry
\begin{subequations}
\begin{align}
    \cal{T}_{\mn}^{\prime} &= e^{2\omega} \cal{T}_{\mn} \ , \\
    \cal{E}^{\prime  \mn} &= e^{2\omega} \cal{E}^{\mn} \ , \\
    \varphi' &= \varphi - 2 \omega \ ,
\end{align}
\end{subequations}
for some function $\omega$ on spacetime.

\printbibliography

\end{document}